\documentclass{aa}  

\usepackage{graphicx,subfigure}
\usepackage{txfonts}
\usepackage[unicode]{hyperref}
\newcommand{\cotan}{\mathrm{cotan}}
\begin{document}

%\title{X-ray analysis of the massive O-type binary HD\,93205}
\title{Investigation of the nature of the wind interaction in HD\,93205 based on multi-epoch X-ray observations}
  % \subtitle{I. Overviewing the $\kappa$-mechanism}

   \author{Bharti Arora
          \inst{1} \and  Micha\"{e}l De Becker\inst{1}%\fnmsep\thanks{Just to show the usage of the elements in the author field}
          }

   \institute{Space Sciences, Technologies and Astrophysics Research (STAR) Institute, University of Li\`ege, Quartier Agora, 19c, All\'ee du 6 A\^out, B5c, B-4000 Sart Tilman, Belgium\\ 
              \email{bhartiarora612@gmail.com}
             }

   \date{Received...; accepted...}
\authorrunning{B. Arora et al.}
\titlerunning{X-ray investigation of the wind interaction in HD 93205}

% \abstract{}{}{}{}{}
% 5 {} token are mandatory
 
  \abstract
  % context heading (optional)
  % {} leave it empty if necessary  
   {The study of the X-ray emission from massive binaries constitutes a relevant approach to investigate shock physics. The case of short period binaries may turn out to be quite challenging, especially in very asymmetric systems where the primary wind may overwhelm that of the secondary in the wind interaction.}
  % aims heading (mandatory)
   {Our objective consists in providing an observational diagnostic of the X-ray behaviour of HD\,93205, that is a very good candidate to investigate these aspects.}
  % methods heading (mandatory)
   {We analysed 31 epochs of \textit{XMM-Newton} X-ray data spanning about two decades to investigate its spectral and timing behaviour. }
  % results heading (mandatory)
   {The X-ray spectrum is very soft along the full orbit, with a luminosity exclusively from the wind interaction region in the range of 2.3 -- 5.4\,$\times$10$^{32}$\,erg\,s$^{-1}$. The light curve peaks close to periastron, with a rather wide pre-periastron low-state coincident with the secondary's body hiding a part of the X-ray emitting region close to its surface. We determined a variability time scale of 6.0807\,$\pm$\,0.0013\,d, in full agreement with the orbital period. Making use of a one-dimensional approach to deal with mutual radiative effects, our results point to a very likely hybrid wind interaction, with a wind-photosphere occurring along most of the orbit, while a brief episode of wind-wind interaction may still develop close to apastron. Beside mutual radiative effects, the radiative nature of the shock that leads to some additional pre-shock obliquitity of the primary wind flow certainly explains the very soft emission.}
  % conclusions heading (optional), leave it empty if necessary
   {HD\,93205 constitutes a relevant target to investigate shock physics in short period, asymmetric massive binary systems, where various mutual radiative effects and radiative shocks concur to display an instructive soft X-ray behaviour. HD\,93205 should be considered as a valid, though challenging target for future three-dimensional modelling initiatives.}

   \maketitle
%
%________________________________________________________________

\section{Introduction}\label{intro}

The investigation of the soft X-ray emission (below 10 keV) from massive (made of O-type and/or Wolf-Rayet stars) binaries constitutes a highly relevant approach to investigate shock physics, providing some insight into stellar winds. In these systems, the strong stellar winds are likely to collide, leading to strong shocks responsible for plasma heating up to several tens of 10$^{6}$\,K. The properties of this thermal X-ray emission depend intimately on the shock parameters, such as pre-shock velocity and wind density. Even though in many cases the wind interaction is basically following a standard behaviour well described by seminal theoretical works \citep{1992ApJ...386..265S,1997MNRAS.292..298P,2010MNRAS.403.1657P}, one can expect some specific cases to happen. This is typically what may happen for very asymmetric systems where one wind is much stronger to the other, potentially leading to wind-photosphere collisions. As a relevant tracer of phenomena occurring in such systems, soft X-ray spectroscopy constitutes a valid tool to characterize their behaviour. The present study focuses on HD\,93205, located in Trumpler\,16, a stellar cluster targeted by several observation campaigns in X-rays over the past decades.  

HD\,93205 is an O-type system displaying a well established double-lined spectroscopic binary nature \citep{1971ApJ...167L..31W,1973ApJ...179..517W}. \citet{1976ApJ...207..502C} classified HD\,93205 as O3V + O8V, and they derived a first orbital solution with a period of 6.08\,d. The orbital solution was refined by \citet{2001MNRAS.326...85M}, pointing to a quite eccentric orbit (see Table\,\ref{hd93205_orb_par1}). HD\,93205 is one of the earliest type star identified in the galaxy having a well constrained orbital solution. The spectral type of the primary star in HD\,93205 was later on revised as O3.5 V((f)) by \citet{2014ApJS..211...10S}. 

Many previous studies reported on the detection of HD\,93205 in X-rays with a moderate X-ray luminosity of $\sim$10$^{32}$ erg s$^{-1}$ (see e.g. \citealt{1979ApJ...234L..55S,1982ApJ...256..530S,1991ApJ...368..241C,1995RMxAC...2...97C}). Clear phase-locked X-ray modulations have been noticed by \citet{1996RMxAC...5...54C} using \textit{ROSAT} observations. Using five X-ray observations from \textit{XMM-Newton}, \citet{2003ASPC..305..383A} have shown that the X-ray flux changes as a function of the separation between two binary components and found the source spectrum to be soft. Some X-ray variability, suspected to be related to colliding winds, have also been reported by \citet{2003MNRAS.346..704A}, \citet{2008A&A...490.1055A}, \citet{2008A&A...477..593A} and \citet{2011ApJS..194....5G}. In spite of the numerous X-ray variability reports previously, a detailed spectral and time analysis of the wind interaction in HD\,93205 is lacking. In particular, the strong asymmetry of the system points to a potentially complex situation, where the primary wind may collide the photosphere, at least at some orbital phases given the eccentricity of this short period system. Making use of a long and consistent time series of X-ray observations, our aim is to characterize its main properties in light of what has been learned over the past decades on the physical circumstances ruling the production of soft X-rays in massive O-type binaries, and especially in short period ones.

This paper is organized as follows. Section \ref{intro} describes the target and summarizes the outcome of previous works relevant to our purpose. Section \ref{x-ray} presents the X-ray observations and the processing of data used for the present study. The X-ray spectral analysis and light curve are detailed in Sections \ref{spec} and \ref{lc}, respectively. The X-ray luminosity of the system is discussed in Sect.\,\ref{sectxlum}. The physical discussion on the origin of the X-ray excess is present in Sect.\,\ref{excess}, before presenting a physically valid scenario for HD\,93205 in Sect.\,\ref{scenario}. Finally, our conclusions are drawn in Section \ref{conc}.

\begin{table}
\centering
	\caption{Main orbital parameters of HD\,93205 as given by \citet{2001MNRAS.326...85M}.}\label{hd93205_orb_par1}
\begin{tabular}{l c}
\hline
\hline
Parameter &  Value \\
\hline
Period (d) & 6.0803 $\pm$ 0.0004 \\
	eccentricity  & 0.370 $\pm$ 0.005 \\	
 $T_{periastron}$ (HJD) & 2450499.089  $\pm$ 0.012  \\
% $V_o$ (km s$^{-1}$) & -2.9 $\pm$ 0.9 \\
 \textbf{$\omega$} (degrees) & 50.8 $\pm$ 0.9 \\
		\bf{Primary:}               &             \\
            $M_{P}$ sin$^{3}$\,\textit{i} ($M_\odot$) & 31.5$\pm$1.1  \\
            %\textit{K}$_{1}$ (km s$^{-1}$) & 132.6 $\pm$ 2.0 \\
            $a_P$ sin\,\textit{i} (cm) & (1.03 $\pm$ 0.02) $\times$ 10$^{12}$ \\
             \bf{Secondary:}               &             \\
			$M_{S}$ sin$^{3}$\,\textit{i} ($M_\odot$) & 13.3$\pm$1.1 \\
   %\textit{K}$_{2}$ (km s$^{-1}$) & 313.6 $\pm$ 1.8 \\
   $a_S$ sin\,\textit{i} (cm) & (2.44 $\pm$ 0.02) $\times$ 10$^{12}$ \\
			\hline
\end{tabular}
\tablefoot{%Here, $V_o$ is the systemic velocity or radial velocity of the center of mass of the system,  \textit{K} is the radial velocity semi-amplitude, 
\textbf{$\omega$} is the longitude of periastron, $T_{periastron}$ is the time of periastron passage, $a$ sin\,\textit{i} is the projected semi-major axis, $M_{P}$ sin$^{3}$\,\textit{i} and $M_{S}$ sin$^{3}$\,\textit{i} are the minimum masses of primary and secondary binary components, respectively.}
\end{table}   
%__________________________________________________________________

\section{X-ray monitoring of HD\,93205 with \textit{XMM-Newton}}\label{x-ray}
X-ray data of HD\,93205 obtained with \textit{XMM$-$Newton} \citep{2001A&A...365L...1J} from 2000 July to 2019 December has been analyzed. A log of total 31 epochs of utilized X-ray observations has been given in Table \ref{tab:log}. These observations were obtained for several other massive stars in Trumpler 16 cluster as main target and HD 93205 was observed as a field star by \textit{XMM$-$Newton} using different configurations of the three European Photon Imaging Camera (EPIC) instruments, \textit{viz.} MOS1, MOS2, and PN. Most of the observations were performed with either thick or medium optical blocking filters. The timing and spectral products have been extracted from data using the latest calibration files with SAS v20.0.0. 

The SAS tasks \textsc{epchain} and \textsc{emchain} were used to process the raw EPIC Observation Data Files (ODF) for PN and MOS observations, respectively. The list of event files were generated considering only the events designated with pattern 0$-$4 for PN and 0$-$12 for MOS data using task \textsc{evselect}. No pileup was noticed in any of the data set upon examining with \textsc{epatplot}. The EPIC background is sometimes affected by high background X-ray photons and/or soft proton flares. In order to eliminate such instances, full-frame light curves were generated considering the single-events (PATTERN = 0) in $>$10 keV energy range for MOS and 10--12 keV band for PN. The events with count rate more than  0.2 counts s$^{-1}$ for MOS and 0.4 counts s$^{-1}$ for PN background light curves were excluded. Rest of the events formed the good time intervals and were utilized for further analysis. 

\begin{figure}[h!]
\centering
    \includegraphics[width=0.98\columnwidth]{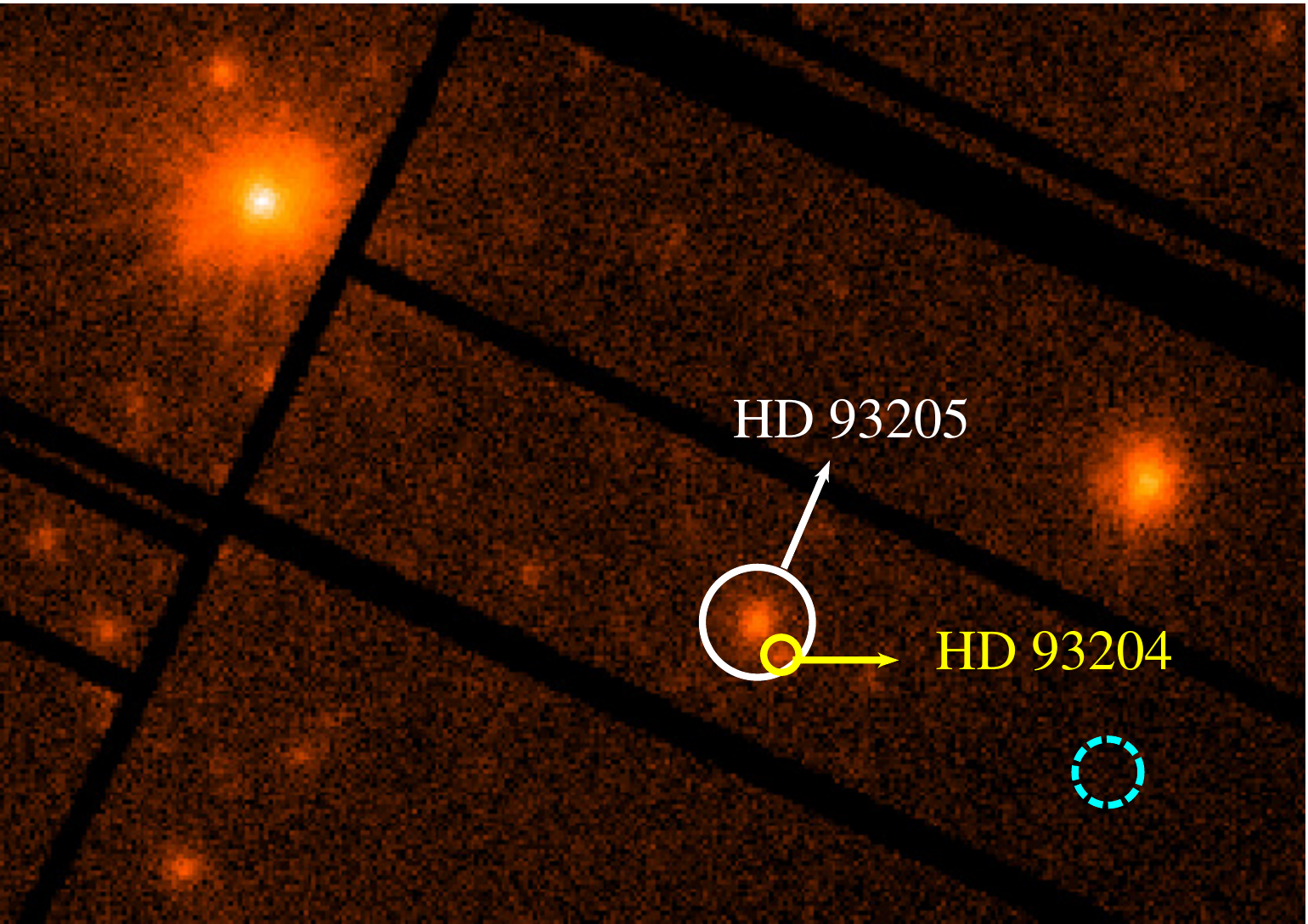}
	\caption{X-ray image of Eta Carinae region extracted from observation ID 0311990101 in 0.3--12 keV energy range. The position of HD 93205 as been highlighted with a white circle of 25 arcsec radius. The X-ray flux of neighboring source HD 93204 has been removed from a circular region of 8 arcsec radius. An estimate of background was done from a circular region (cyan color) having radius 15 arcsec lying in the source-free area of the detector.
 \label{fig:fig1}}
\end{figure}

Fig.\,\ref{fig:fig1} shows the X-ray image extracted from the observation ID 0311990101 in 0.3$-$12.0 keV energy range. The source co-ordinates were precisely determined using the source detection algorithm in SAS named as \textsc{edetect\_chain}. The location of HD 93205 has been detected at the co-ordinates R.A. (J2000) = 10:44:33.84 and Dec (J2000) = -59:44:13.92. HD 93205 has a neighboring source HD 93204 detected as close as $\sim$18.3 arcsec at the position R.A. (J2000) = 10:44:32.40 and Dec (J2000) = -59:44:28.68  as shown in Figure \ref{fig:fig1}.  HD 93204 is an O5.5 V((f)) star emitting in X-rays \citep{2014ApJS..211...10S}. We compared the location of both of these sources with the co-ordinates determined from the high resolution \textit{Chandra} Carina catalog \citep{2011ApJS..194....2B}. Both the positions matched very well confirming our source detection approach. A circular region centered at the estimated location of HD\,93205 with radius 25 arcsec has been selected to extract the light curves and spectra of the target source from EPIC CCDs. HD 93204 lies with in the chosen circular region around HD 93205, therefore, the counts from another circular region of radius 8 arcsec centered at the location of HD 93204 were subtracted from the actual source region. This procedure helped to remove the X-ray contamination from the neighboring source to HD 93205 X-ray products.    

The background light curve and spectrum were extracted from a 15 arcsec radius circular region located at the source-free positions surrounding HD 93205. The proximity of background region to the main target ensures roughly same off-axis angle of these two regions on the CCD plane. \textit{Chandra} images were also investigated to verify that no other X-ray emitting object is present in our chosen background region due to its better resolution and sensitivity than XMM-Newton--EPIC. In order to make corrections in the extracted light curves to consider the good time intervals, dead time, exposure, point-spread function, and background subtraction,  the task  \textsc{epiclccorr} has been used. The source as well as the background spectra were generated by \textsc{evselect} task.  \textsc{arfgen} and \textsc{rmfgen} were employed to calculate the dedicated Ancillary Response Files (ARFs) and Response Matrix Files (RMFs), respectively, necessary for the spectral fitting. The extracted spectra were backscaled using the task \textsc{backscale}. The energy channels in each of the extracted  source spectrum were grouped using \textsc{grppha} to have a minimum of 10 counts per spectral bin. Further temporal and spectral analyses were performed using HEASoft version v6.29c.
                                                                                              
%______________________________________________________________

\section{X-ray spectral analysis}\label{spec}
\begin{figure}[h!]
\centering
 \includegraphics[width=1.0\columnwidth,trim={0.0cm 0.0cm 4.0cm 1.0cm}]{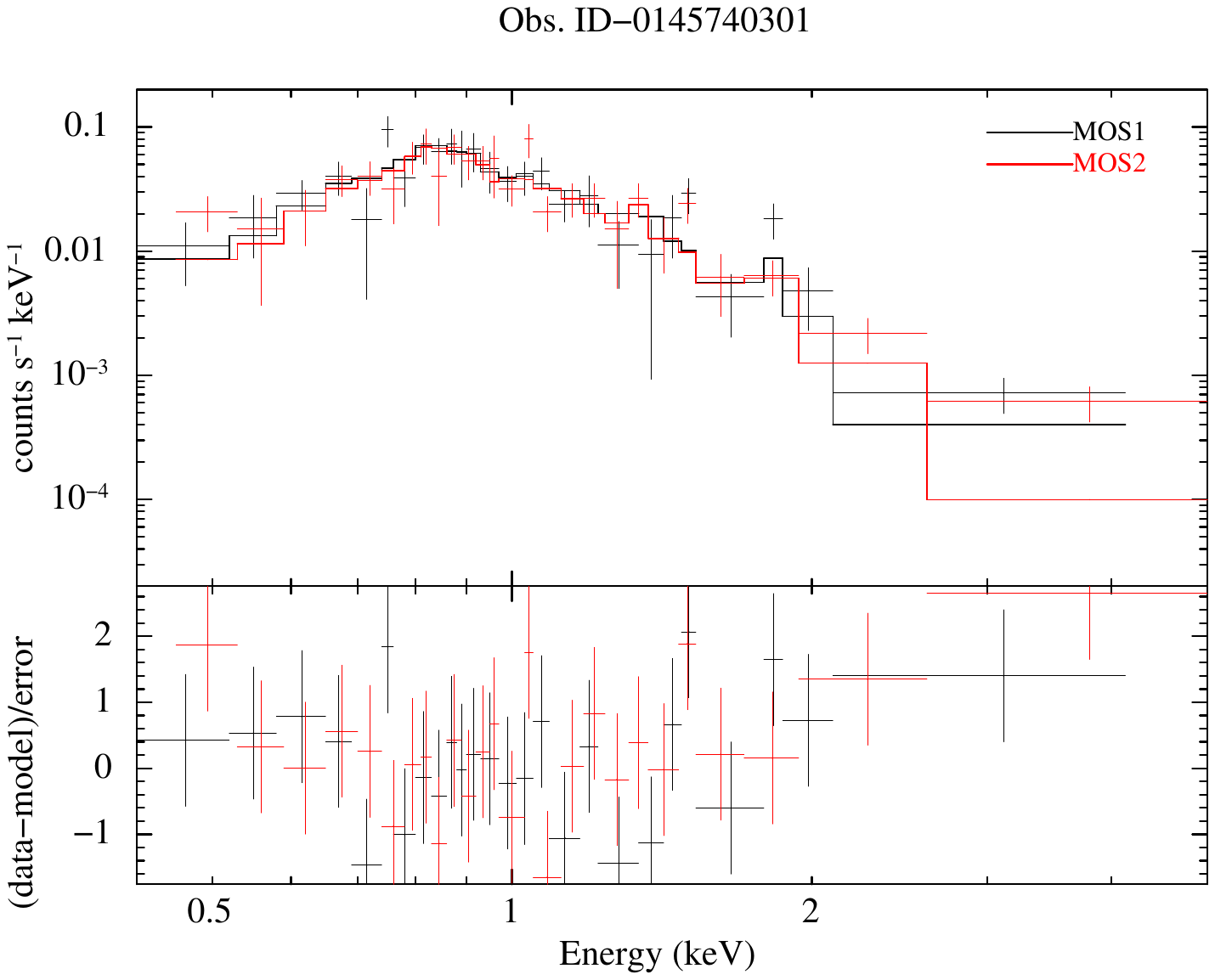}
\caption{MOS1 and MOS2 spectra of HD\,93205 jointly fitted with the two-temperature thermal plasma emission model from the observation ID 0145740301. The details of the spectral fitting are provided in Table \ref{spec_par}.
\label{fig:hd93205_spectra}}
\end{figure}

The X-ray spectra of HD 93205 extracted consistently from \textit{XMM-Newton}--EPIC displays typical features of an optically thin plasma heated in hot stars' winds above 10$^{6}$ K. The obtained spectra were relatively soft with maximum emission located between 0.8$-$0.9 keV as shown in Fig. \ref{fig:hd93205_spectra}. There were no significant counts with energy more than 5 keV above background level. Therefore, the further spectral energy analysis was restricted to 0.3--5.0 keV energy range.  

In order to characterize the X-ray spectral properties of HD\,93205, and search for any variability with time, we considered optically thin thermal emission models. Such an approach is standard for the modelling of the X-ray emission from massive binaries \citep{2010MNRAS.403.1657P,2015MNRAS.451.1070D,2019MNRAS.487.2624A,2021AJ....162..257A,2024A&A...687A..34A}. We made use of the {\sc apec} emission model available in the X-ray spectral fitting package XSPEC (v.12.12.0). We fitted the spectrum with multiple {\sc apec} components but using only two emission components turned out to be sufficient to model the X-ray spectrum of HD\,93205. The composite model was of the form \textsc{phabs(ism)*phabs(local)*(apec+apec)}. The multiplicative photoelectric absorption component {\sc phabs} has been used to model the effects of interstellar absorption and local absorption by the stellar wind material on the emitted X-rays. The interstellar column of absorbing matter was fixed to value of $N_{H}^{ism}=0.24\times10^{22}$ cm $^{-2}$ obtained from \citet{2019ApJ...872...55J}. However, local column ($N_{H}^{local}$) was kept as a free parameter during the fitting procedure. The reduced $\chi^2$ minimization was used to test the goodness of fit. The model assumed solar abundances from \citet{1989GeCoA..53..197A} while performing the spectral fitting. 

Independent modelling of the spectra obtained at different epochs is not straightforward, as it may lead to a significant dispersion in the best-fit parameters, depending upon the quality of individual data-sets. The PN spectra has more number of counts than other EPIC detectors, therefore, we fitted only PN spectra initially with the two-temperature composite model to explore the parameter space and achieve a first diagnostic of the full time series, without freezing any parameter (apart from interstellar absorption). Upon modelling the PN spectra obtained at each epoch separately, it was found that the average values of two temperatures corresponding to soft and hard emission were 0.16$\pm$0.01 (=$kT_{1}$) and 0.60$\pm$0.02 (=$kT_{2}$), respectively. The achieved value of reduced $\chi^2$ was below 1.3 for all of the data sets. However, for few of the spectra the temperature corresponding to hard emission fitted at higher value around 1.5 keV which better modelled the high energy tail but overall it resulted in poor reduced $\chi^2$ than low $kT_{2}$ model and it resulted in increased normalisation constant of soft component ($norm_1$) at the expense of normalisation value from hard component ($norm_2$). As it did not improve the physical relevance of the spectral fitting, we decided to proceed further with $kT_{1}$=0.2 keV and $kT_{2}$=0.6 keV. These values are consistent with the ones estimated by \citet{2008A&A...477..593A} and \citet{2020RAA....20..108R}. 

We note that similar results were obtained for MOS1 and MOS2 spectra, so we decided to focus on the joint fitting of all EPIC spectra at each epoch. As the next step, we decided to freeze some parameters to constrain the exploration of the parameter space and achieve a consistent set of solutions for the full-time series. All emission model temperatures were frozen ($k$T$_{1}$ = 0.2 keV and $k$T$_{2}$ = 0.6 keV), keeping the local absorption column and the two normalization parameters free. Following this approach, we obtained a consistent modeling series for 31 epochs, including cases where the number of counts were too low to obtain adequate results in our first unconstrained attempts. All our results are presented in Table\,\ref{spec_par}. Reduced $\chi^2$ values range between 0.9 and 1.5. 

\begin{figure}[h!]
\centering
\includegraphics[width=0.99\columnwidth,trim={0.0cm 1.0cm 0.0cm 0.0cm}]{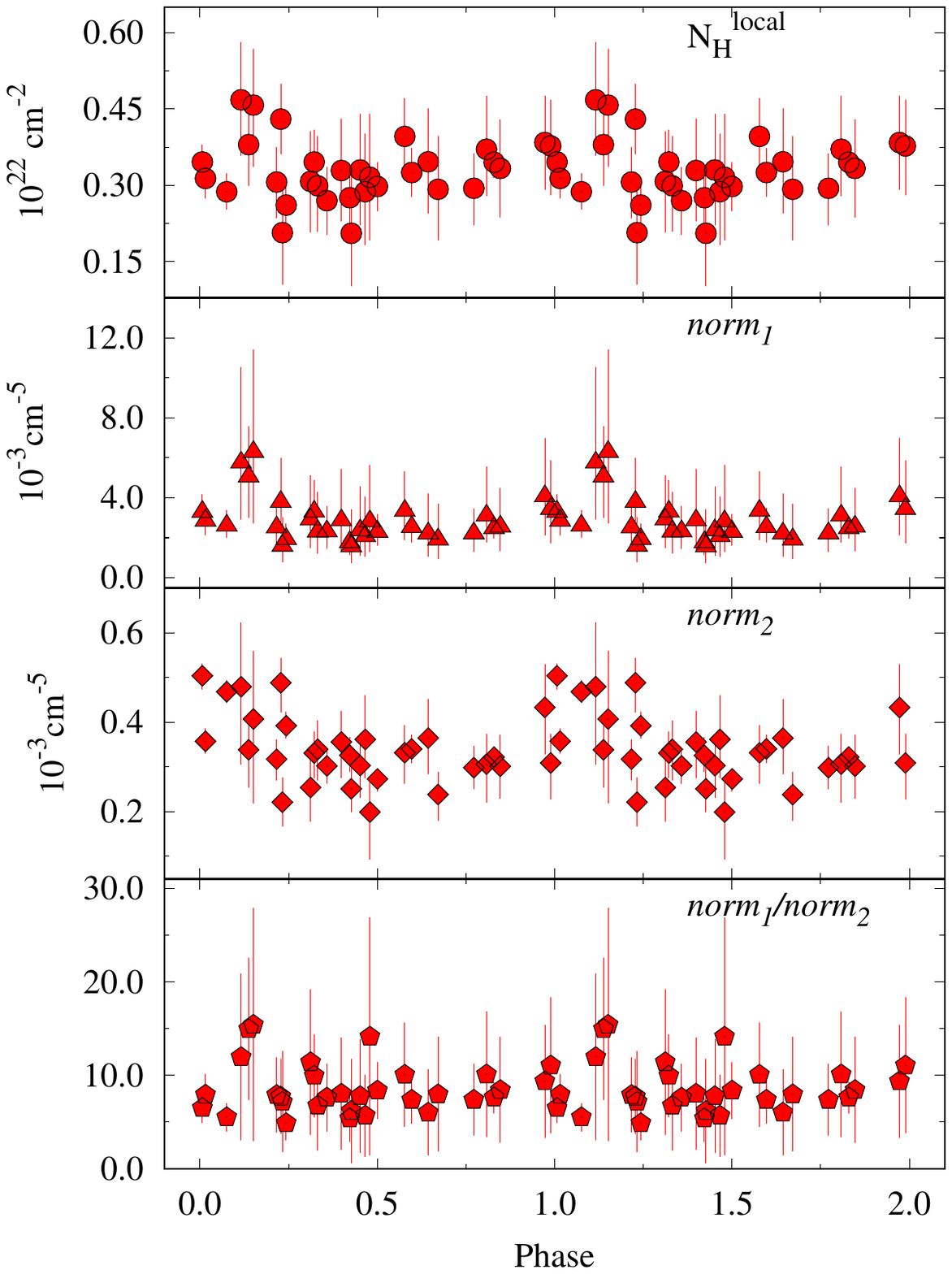}
%\vspace*{-0.7cm}
\caption{Variation of the local equivalent H-column density (N$_{H}^{local}$), normalization constants ($norm_{1}$ and $norm_{2}$) and the ratio $norm_{1}$$/$$norm_{2}$ corresponding to two thermal plasma emission components with the orbital phase of HD\,93205 (see Table \ref{spec_par})}.
\label{fig:hd93205_spec_par}
\end{figure}

\begin{figure*}
\centering  
\subfigure[Observed flux]{\includegraphics[width=0.97\columnwidth,trim={0.4cm 1.5cm 0.0cm 5.0cm}]{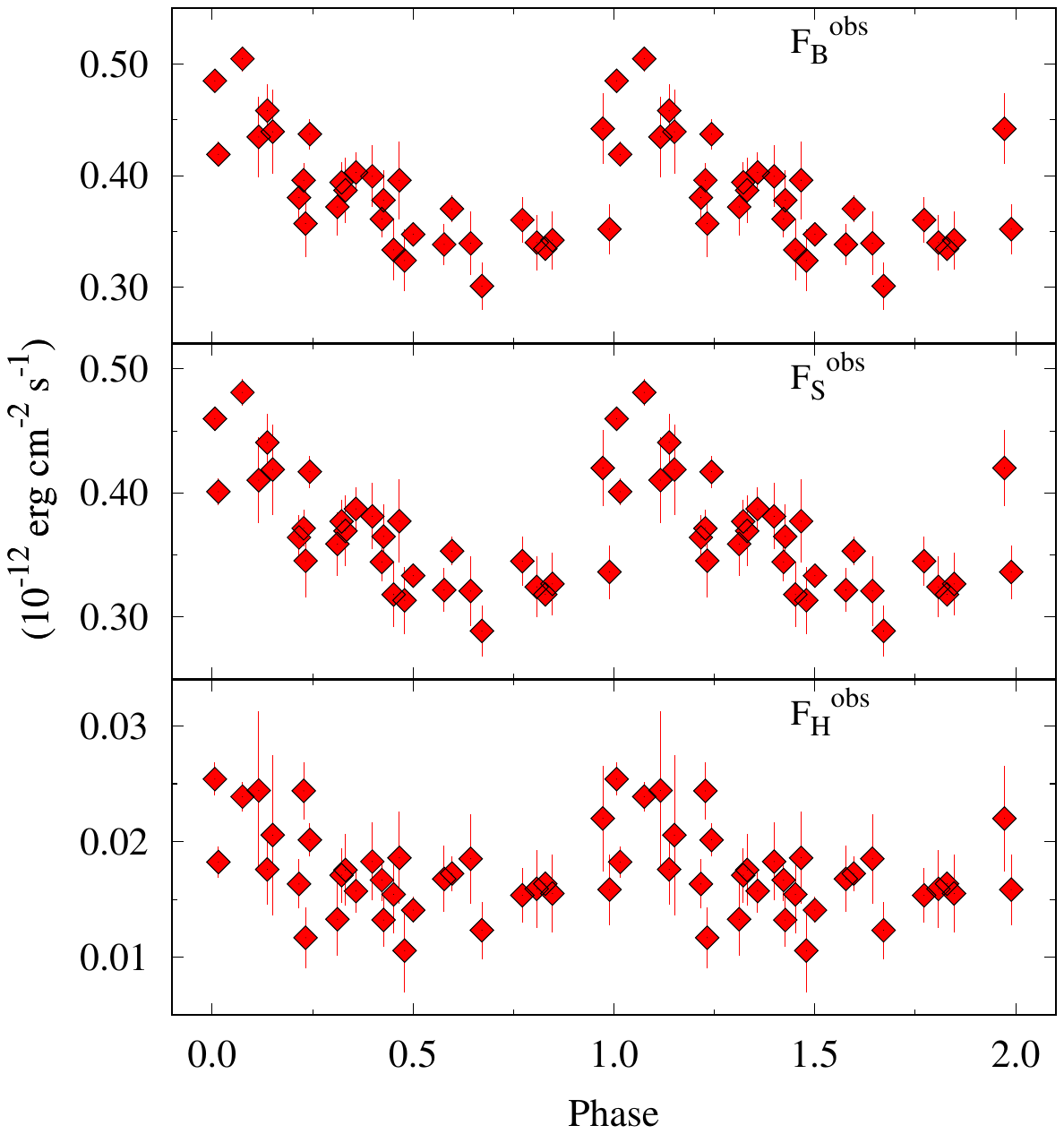}}
\subfigure[ISM corrected flux]{\includegraphics[width=0.97\columnwidth,trim={0.4cm 1.5cm 0.0cm 5.0cm}]{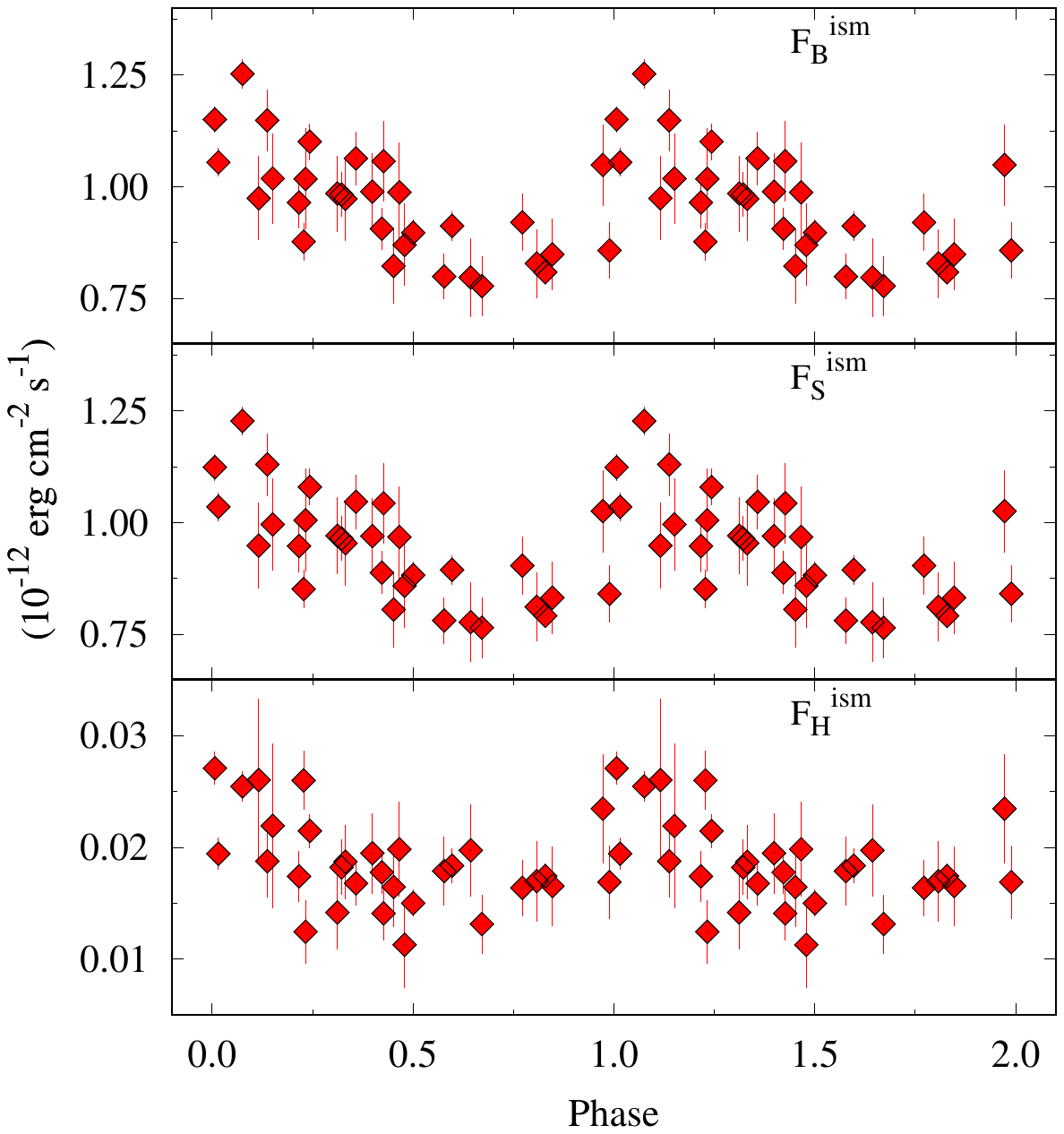}}
%\vspace*{-0.9cm}
\caption{Modulation of the (a) observed ($F^{obs}$) and (b) ISM-corrected ($F^{ism}$) X-ray flux in broad, soft, and hard energy bands from HD\,93205 obtained after X-ray spectral fitting as a function of the binary orbital phase (see Table \ref{spec_par}).
\label{fig:flux_phase}}
\end{figure*}

The MOS1 and MOS2 X-ray spectra of HD 93205 shown in Fig.\,\ref{fig:hd93205_spectra} has been jointly modelled. In order to see the evolution of various spectral parameters over the complete orbit of HD 93205, we estimated its binary orbital phase according to the ephemeris provided by \citet{2001MNRAS.326...85M} as HJD $=$ 2450499.089$+$6.080E. Zero phase corresponds to the time of periastron passage in this eccentric binary system. The best-fit parameters $N_H^{local}$, $norm_1$ and $norm_2$ have been phase phase-folded in Fig.,\ref{fig:hd93205_spec_par}. The ratio $norm_1$/$norm_2$  plotted in the bottom panel of this figure provides hints about variation in hardness of spectrum at several epochs of observation. The observed and ISM-corrected X-ray flux was also estimated in 0.3--5.0 keV (broad, F$_{B}^{obs,ism}$), 0.3--2.0 keV (soft, F$_{S}^{obs,ism}$), and 2.0--5.0 keV (hard, F$_{H}^{obs,ism}$) energy bands. The variation of X-ray flux as a function of the orbital phase is shown in Fig.\,\ref{fig:flux_phase} in different energy bands. The observed trend suggests that in the broad and soft energy band there is a variation in flux by about a factor of $\sim$1.6 between its lower and higher value achieved after apastron and close to periastron, respectively. However, in hard energy band the ratio is $\sim$2.4. The phase-locked X-ray flux variations are clearly in agreement with the modulations shown in the Fig.\,\ref{fig:hd93205_spec_par}. 

\section{X-ray timing analysis}\label{lc}

\begin{figure}
\centering
  \includegraphics[width=0.98\columnwidth,trim={0.0cm 1.0cm 0.0cm 4.5cm}]{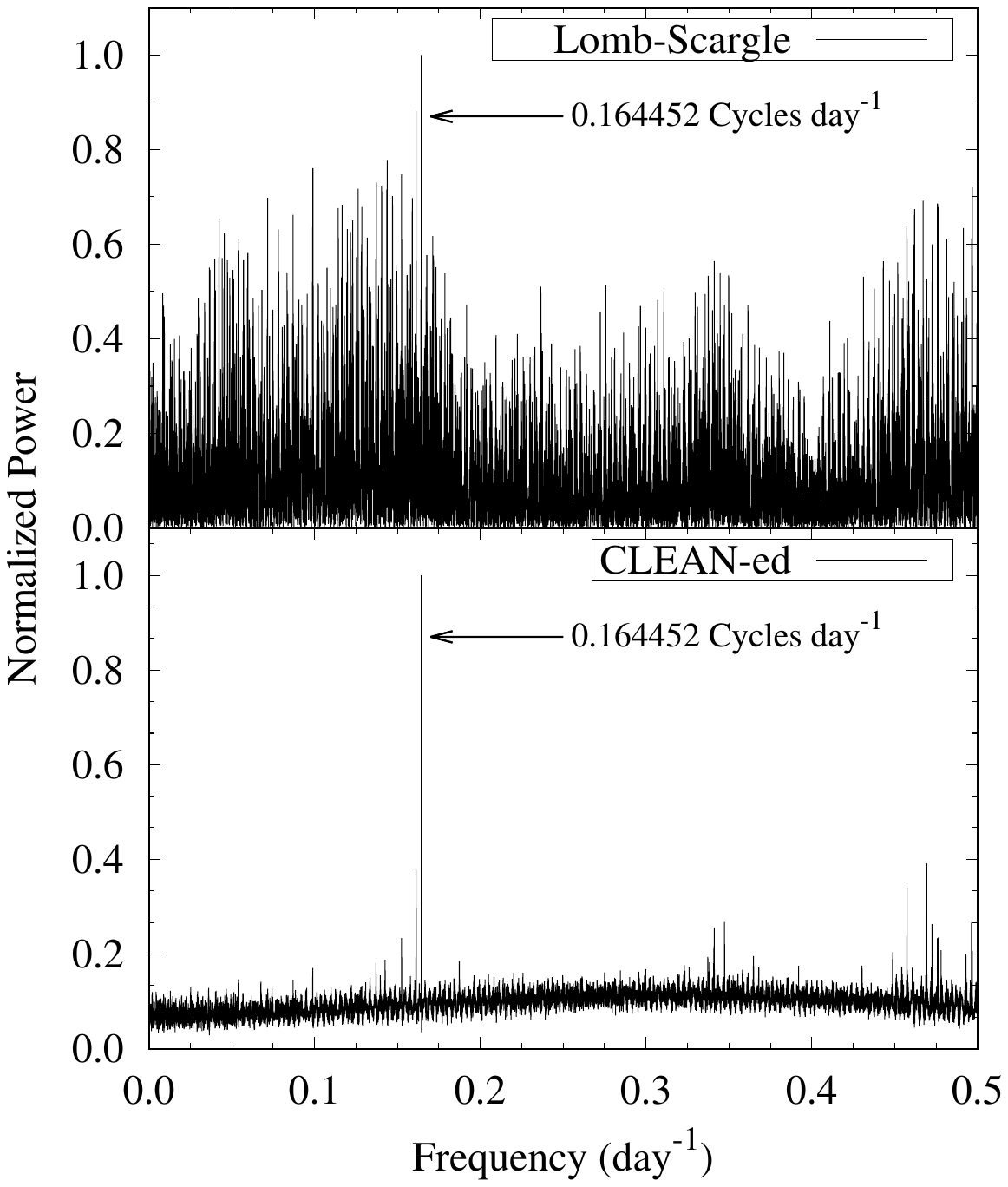}
%\vspace*{-0.7cm}
\caption{Lomb-Scargle and CLEANed power spectra of observed flux ($F^{obs}_{B}$) from HD\,93205 in the 0.3--5.0 keV energy range. The frequency of the peak with the highest power is also mentioned.
\label{periodogram}}
\end{figure}

In order to verify the determination of orbital period of HD 93205, we have performed a Fourier transform (FT) of the flux measurements provided in Table\,\ref{spec_par}. We used both Lomb-Scargle \citep{1976Ap&SS..39..447L,1982ApJ...263..835S,1986ApJ...302..757H} and CLEAN \citep{1987AJ.....93..968R} periodograms. Both algorithms were applied on the flux values obtained at 31 epochs. The Lomb-Scargle algorithm is effective in searching for periodic signals in unevenly time sampled light curves and as shown in upper panel of Fig. \ref{periodogram}. It consists of several peaks along with one dominant peak at frequency 0.16445$\pm$0.00004 cycles d$^{-1}$. In order to remove several other peaks which might be originating due to the aliasing, the CLEAN power spectrum was also calculated with a loop-gain of 0.1 and 100 iterations. The CLEANed periodogram showed a clear, distinguished peak at exactly the same frequency as of the Lomb-Scargle power spectrum in the bottom panel of Fig. \ref{periodogram}. The peak corresponds to a period of 6.0807$\pm$0.0013 days in full consistency with the orbital period obtained by \citet{2001MNRAS.326...85M} through high-quality spectroscopic solution. The present period determination shows relevance of investigating massive stars multiplicity from X-ray time series analysis carried over a considerably long time baseline to the classical orbit determination techniques (e.g. see \citealt{2024A&A...687A..34A}). 

%______________________________________________________________

\begin{figure*}
\centering
\subfigure[MOS1]{  \includegraphics[width=0.47\textwidth,trim={0.0cm 1.0cm 0.0cm 5.0cm}]{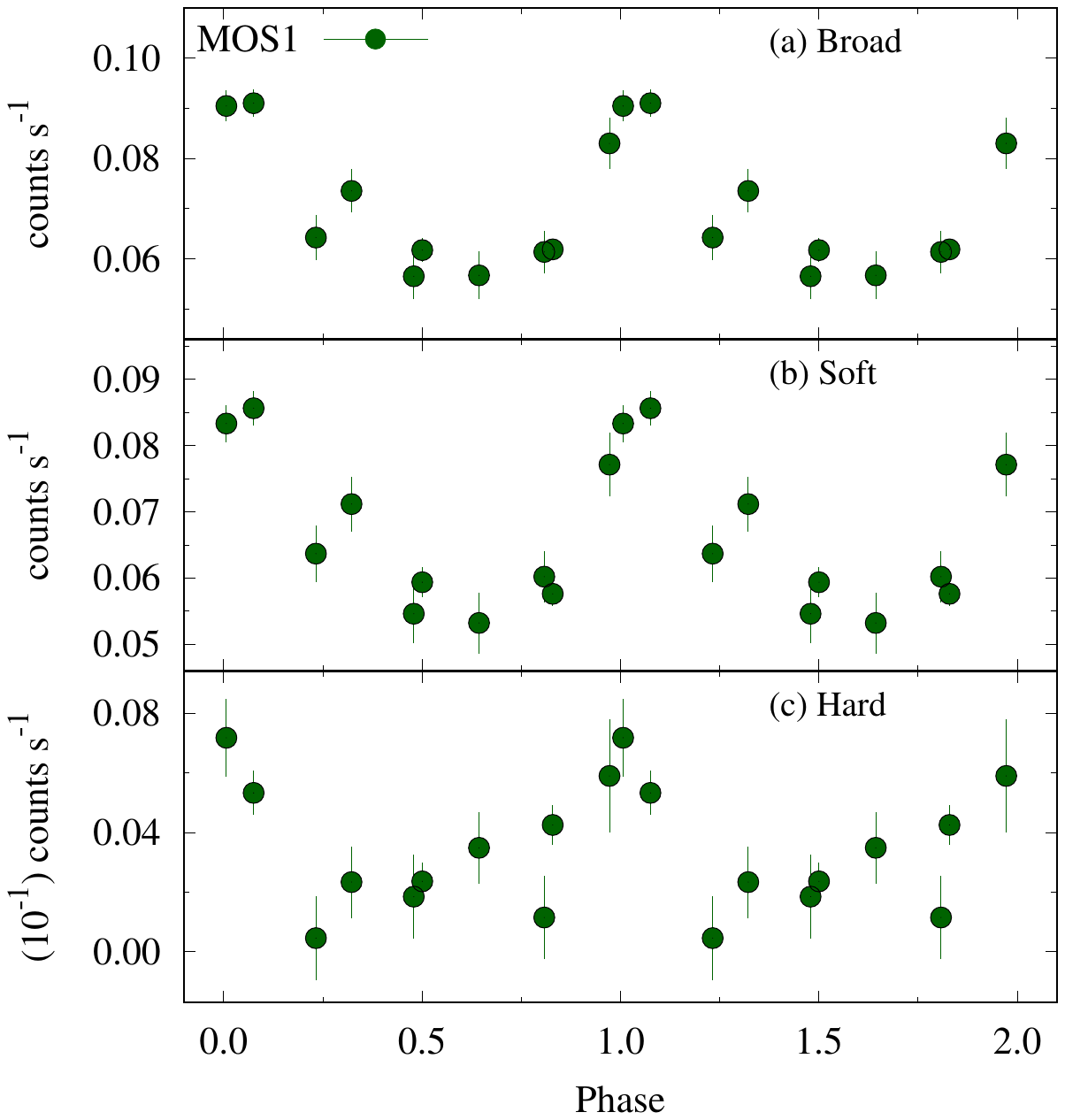}}
\subfigure[MOS2]{  \includegraphics[width=0.47\textwidth,trim={0.0cm 1.0cm 0.0cm 5.0cm}]{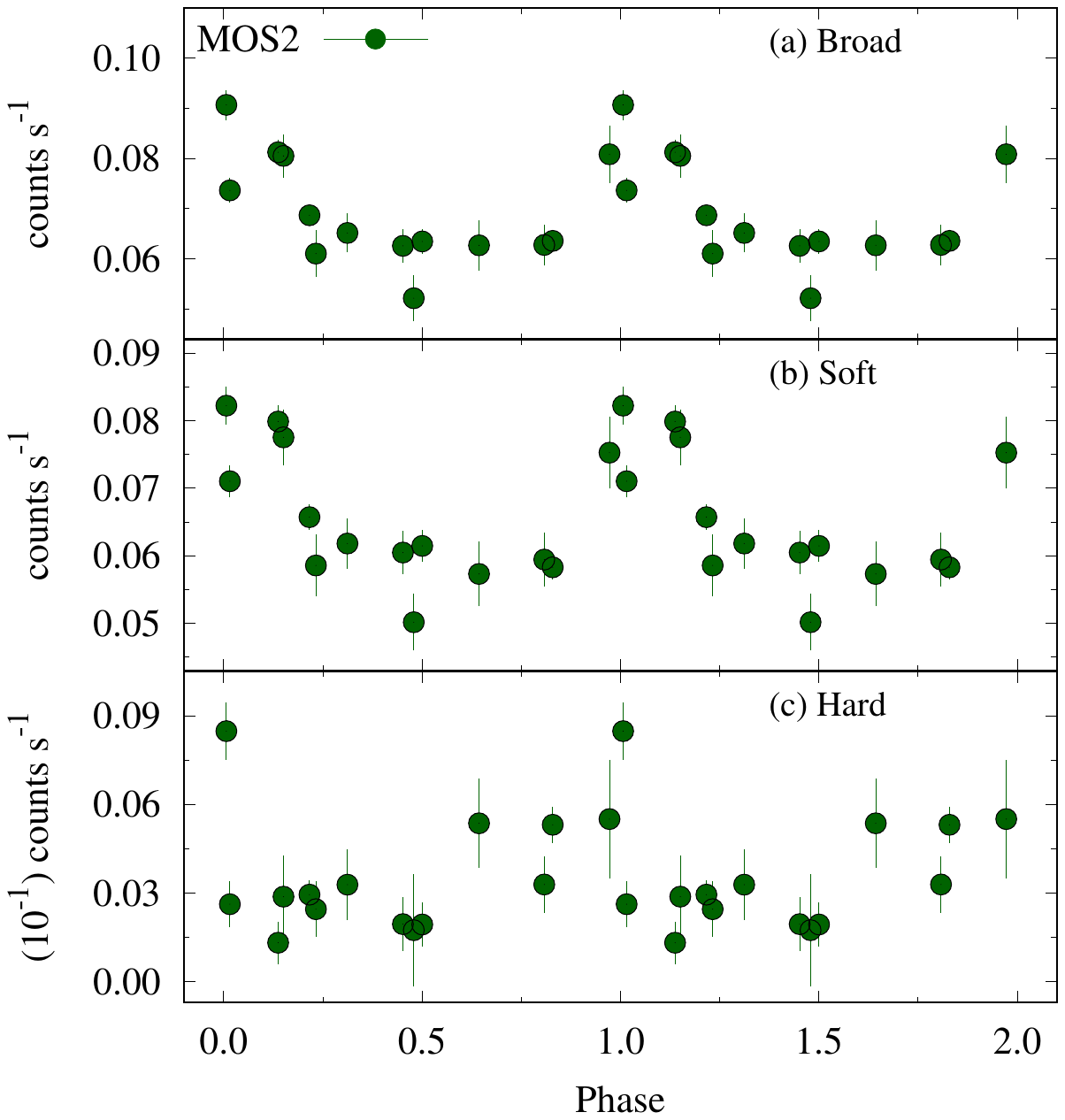}}
\caption{Folded X-ray light curves of HD\,93205 as observed by \textit{XMM-Newton} using (a) MOS1, and (b) MOS2 in broad, soft, and hard energy bands. The average count rate of each individual light curve observed with thick optical blocking filter has been plotted here.
\label{fig:hd93205_flc}}
\end{figure*}

HD\,93205 was properly exposed mainly in MOS1 and MOS2, therefore providing us with adequate time series for light curve analysis. The background-subtracted X-ray light curves as obtained from MOS1 and MOS2 are shown in Fig. \ref{fig:hd93205_flc}. These light curves were extracted in broad, soft, and hard energy bands. While plotting these light curves, we have considered the count rates estimated from the observations obtained with thick optical blocking filters as given in table \ref{tab:log}. It has to be noted that the difference in the detector effective area is significant below 0.6--0.7 keV for MOS\footnote{\url https://xmm-tools.cosmos.esa.int/external/xmm\_user\_support/\\documentation/uhb/epicfilters.html}. As noticed in Section \ref{spec}, most of the X-ray emission from HD 93205 comes from the soft energy range therefore it becomes important to account for detector efficiency while analysing variation in raw count rate obtained from X-ray light curves. We could see clear distinction in the count rates taken with different filters for the same instrument. However, the detector effective area is taken into picture by using dedicated ARF files while spectral fitting, facilitating analysis of all data points together in Fig. \ref{fig:flux_phase}. The light curves in Fig. \ref{fig:hd93205_flc} were orbital phase folded and each point corresponds to average count rate from individual observation ID. 

The X-ray variability is seen in all the energy band light curves with an order of magnitude lower count rate in the hard band. The phase-locked modulation in count rate is almost similar in the broad and soft energy bands where it becomes maximum around orbital phase zero and drops gradually to minimum around apastron. This behaviour is more systematic in broad and soft energy bands. The count rate enhanced by $\sim$60\% in these energy band from apastron to periastron, although variation in hard band is more than 200\% for MOS1 light curves which has clearer variation than MOS2 light curves. Finally, it is emphasised that soft band accounts for more than 90\% of the total counts detected in broad energy band in good agreement with the soft X-ray spectrum discussed in Sect.\,\ref{spec}.

%______________________________________________________________

\section{X-ray luminosity}\label{sectxlum}
The fluxes corrected for interstellar absorption were converted to X-ray luminosities, assuming a distance of 2350 pc \citep{2021AJ....161..147B}. Given the flux values across the orbit quoted in Table\,\ref{spec_par}, we determined luminosities (L$_{X}$) ranging between 5.2\,$\times$\,10$^{32}$ and 8.3\,$\times$\,10$^{32}$ erg\,s$^{-1}$. 

A comparison with the X-ray luminosity reported by \citet{2011ApJS..194....5G} shows that their value ($\sim 3.6\,\times\,10^{32}$\,erg\,s$^{-1}$) is lower than the luminosity range reported on above by about a factor 1.5. Their {\it Chandra} measurement spanned an orbital phase range (0.21--0.38) away from the maximum of the light curve. We attribute this discrepancy to two reasons. First, \citet{2011ApJS..194....5G} worked in the 0.5--8.0 keV energy range, while we used the 0.3--5.0 keV band. Since most of the X-ray emission from HD\,93205 is soft, our estimate of the X-ray luminosity accounts for a non-negligible contribution that was not measured by \citet{2011ApJS..194....5G}. Second, our modelling points to a major contribution from a plasma with a temperature of about 0.2 keV with a weak contribution from a plasma at 0.6 keV (versus 0.23 and 0.74 keV for \citealt{2011ApJS..194....5G}). Such a spectrum is especially bright in the low energy part of the X-ray spectrum that is also the most affected by interstellar absorption. Because of the slightly softer spectrum considered in our study, the correction for that interstellar absorption is more pronounced in our case, leading to ISM-corrected luminosities somewhat higher than reported by \citet{2011ApJS..194....5G}.

\begin{table}[h!]
\caption{Adopted parameters for both components of HD\,93205, based on \citet{2005A&A...436.1049M} and \citet{2012A&A...537A..37M}.}\label{param}
\centering  
\begin{tabular}{l c c}
\hline
 & O3V & O8V \\
\hline
$L_{bol}$ (erg\,s$^{-1}$) & $2.60\,\times\,10^{39}$ & $0.31\,\times\,10^{39}$ \\
${\dot M}$ (M$_\odot$\,yr$^{-1}$) & $2.29\,\times\,10^{-6}$ & $9.6\,\times\,10^{-8}$\\
$V_\infty$ (km\,s$^{-1}$) & 3790 & 2450 \\
$T_{eff}$ (K) & 44600 & 33400 \\
$R$ (R$_\odot$) & 13.84 & 8.52 \\
\hline
\end{tabular}
\tablefoot{For the secondary, the mass loss rate is based on \citet{2001A&A...369..574V}, as values from \citet{2012A&A...537A..37M} for late O dwarfs deserve some caution. Index $P$ and $S$ will be used to identify parameters from the primary and secondary, respectively.}
\end{table}

For O-type stars, the intrinsic X-ray emission from individual stellar winds is expected to be about 10$^{-7}$ times the bolometric luminosity (L$_{bol}$). This luminosity ratio is valid for not too evolved O-type stars \citep{2013MNRAS.429.3379O,2013NewA...25....7D}. Given the values of parameters quoted in Table\,\ref{param}, the measured X-ray luminosity points to an X-ray over-luminosity factor in the range of 1.8 -- 2.8. This translates into an excess luminosity $L_{X,excess} = L_{X} -  L_{X,P} - L_{X,S}$ = 2.3 -- 5.4\,$\times$\,10$^{32}$ erg\,s$^{-1}$ (where $L_{X,P}$ and $L_{X,S}$ are the individual X-ray luminosities from the primary and secondary winds, respectively). For the O3 primary, the thick wind may be opaque enough to lead to the luminosity ratio to be slightly lower than 10$^{-7}$. Our estimate of L$_{X,excess}$ may thus be seen as a conservative lower limit. This excess emission produced on top of individual winds constitutes the core of our discussion in Sect.\,\ref{excess}.

\section{Origin of the X-ray emission excess}\label{excess}

\subsection{Wind interaction}\label{windinter}
In a massive binary system, a significant excess X-ray emission is usually attributable to colliding winds \citep{1992ApJ...386..265S,2010MNRAS.403.1657P}. However, given the highly asymmetric nature of HD 93205, the primary wind may completely overwhelm the secondary one, therefore preventing such a wind-wind collision (WWC) to occur. This opens up the possibility for a wind-photosphere collision (WPC) to develop in the system.

We searched for stable solutions of the stagnation point on the basis of the ram pressure balance condition. Given the small size of the orbit, stellar winds do not reach their terminal velocity before colliding. We assumed a classical $\beta$-velocity law (with $\beta = 0.8$)
\begin{equation}\label{betalaw}
V = V_\infty\,\bigg(1 - \frac{R}{D}\bigg)^\beta    
\end{equation}
for both winds (where $R$ and $D$ are the stellar radius and distance available to the wind to be radiatively accelerated before colliding, respectively), along with wind parameters quoted in Table\,\ref{param}. For the evolution of the stellar separation along the orbit, we assumed parameters taken from Table\,\ref{hd93205_orb_par1}. By a comparison between minimum masses and expected absolute masses given by \citet{2005A&A...436.1049M}, we deduce an inclination of about 58$^\circ$ (averaging values from both stars), in fair agreement with the 55$^\circ$ suggested by \citet{2001MNRAS.326...85M}. This results in an absolute separation of 58.8\,R$_\odot$, that has to be modulated by the eccentricity depending on the orbital phase. Our analysis resulted in a lack of stable stagnation point between the two stars at any orbital phase, pointing to the very likely scenario of a collision between the primary's (stronger) wind and the secondary's photosphere. However, we note that different assumptions on the wind parameters (including in particular a downward revision of the primary mass loss rate) may allow a stable WWC to form quite close to the secondary's surface. This deserves to be qualified by a more detailed discussion including the influence of radiative effects. 

\subsection{Influence of radiative effects}\label{radeffect}
The dynamics of the incoming primary wind is very likely to be affected by radiative effects. Fist of all, in strongly asymmetric systems such as HD\,93205, one may be dealing with {\it sudden radiative braking} \citep{1997ApJ...475..786G}. This process considers that the primary wind is suddenly decelerating while getting very close to the secondary's surface. This may prevent the WPC to occur, leading alternatively to a WWC close to the secondary's surface. The study by \citet{1997ApJ...475..786G} consists in a one-dimensional approach that therefore ignores orbital effects and assumes axial symmetry about the line of separation between the stars. Such a symmetry would be broken in case of significant orbital distortion that tends to increase as the orbital period decreases. A valid indicator of this would be the orbital-to-wind velocity ratio: the greater the ratio, the stronger is the departure from axial symmetry. Despite of this limitation, this approach has the benefit to provide us with some analytical tools to investigate semi-quantitatively the influence of this radiative effect on our system at every orbital phase. A full three-dimensional analysis is well beyond the scope of the the present study.

In order to check for the relevance of the WPC and WWC scenarios, we made use of the formalism presented by \citet{1997ApJ...475..786G}. Unless specified differently, all parameters in equations below are expressed in cgs units. Let's first define the braking coefficient ($C_b$),
\begin{equation}
C_b = \bigg(\frac{V_P}{v_{th}}\bigg)\,\bigg(\frac{L_S\,\kappa_e}{2\,\pi\,V_P^2\,c\,D}\bigg)^{1 - \alpha}\,P_{P/\nu}^{-\alpha}  
\end{equation}
\noindent where $V_P$ is the primary wind velocity, $v_{th}$ is the plasma thermal velocity, $L_S$ is the secondary bolometric luminosity, $\kappa_{e}$ is the free electron cross section per gram, $D$ is the stellar separation, $\alpha$ is the index of the line list power distribution of the standard CAK wind theory \citep{1975ApJ...195..157C}. $P_{P/\nu}$ is the ratio of the primary wind momentum flux to the secondary radiative momentum flux (${\dot M}_P\,c\,V_P/L_S$). 

\begin{figure}
\centering
    \includegraphics[width=1.0\columnwidth]{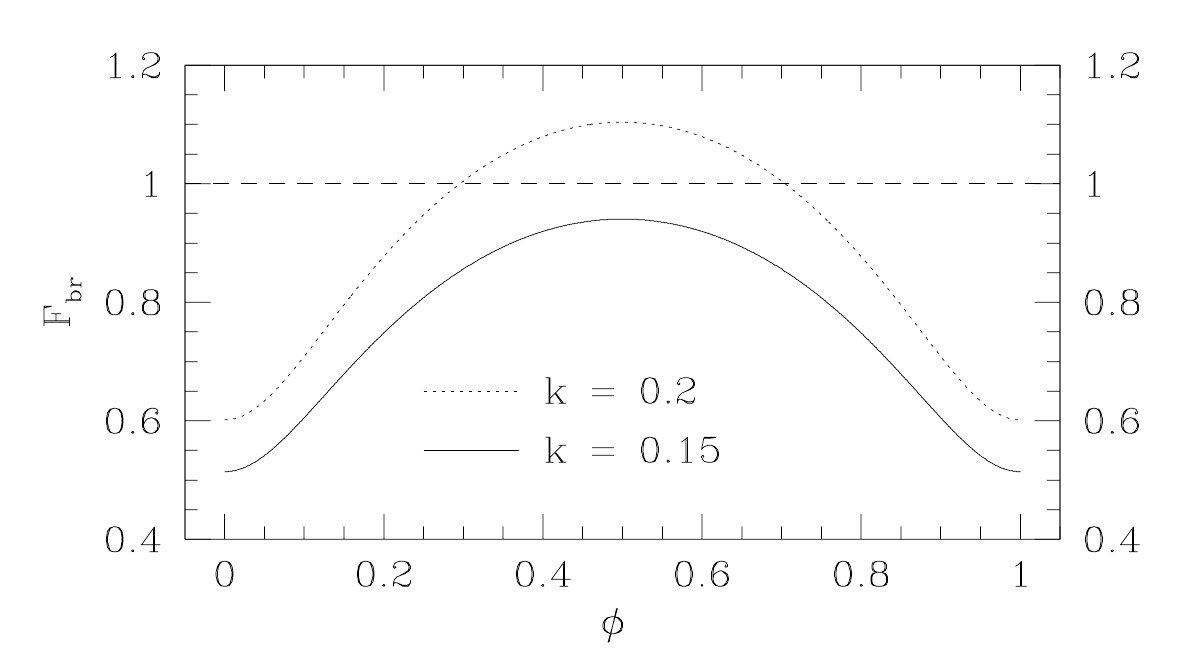}
	\caption{Plot of the braking parameter ($F_{br}$) as a function of the orbital phase. The solid and dotted line stand for two different assumptions for the $k$ parameter. The horizontal dashed line stands for the critical value above which the WPC scenario is not valid.\label{radbr}}
\end{figure}

The so-called braking radius ($x_b$) is expressed as follows,
\begin{equation}
x_b = \Bigg[1 + \bigg(\frac{d}{\eta}\bigg)^{(1 - \alpha)/(1 + \alpha)}\Bigg]^{-1}
\end{equation}
with
\begin{equation}
\eta = \frac{1 - \alpha}{1 + \alpha}\,\bigg(\frac{1}{1 + \alpha}\,k\,C_b\bigg)^{1/(1 - \alpha)}\,d
\end{equation}
\noindent where $k$ is the force multiplier parameter from the CAK theory, and $d$ is the stellar separation expressed in units of the secondary's stellar radius ($d = D/R_S$).

According to \citet{1997ApJ...475..786G}, radiative braking cannot prevent photospheric collision provided the following condition is fulfilled,
\begin{equation}\label{radbr-eqn}
F_{br} = x_b\,d < 1
\end{equation}
\noindent where $F_{br}$ will be referred to as the braking parameter. Adopting $\alpha = 0.6$ and $k = 0.15$ (values valid for O-type stars adopted by \citealt{1997ApJ...475..786G}), along with $\kappa_e = 0.37 g^{-1}$ (not specified by \citealt{1997ApJ...475..786G}, but determined from their results), and making use of parameters quoted in Tables\,\ref{hd93205_orb_par1} ad \ref{param}, we checked for the validity of the condition expressed by Eq.\,\ref{radbr-eqn} along the full orbit. As shown in Fig.\,\ref{radbr}, the braking parameter is lower than one over the full orbit (solid line). In order to check for the robustness of this result, we repeated the calculation adopting other values for $k$ and $\alpha$. $F_{br}$ tends to go above 1.0 when $\alpha$ goes above 0.6, especially close to apastron. For $k = 0.2$, $F_{br}$ is above 1.0 between orbital phases 0.3 and 0.7 (dashed line in Fig.\,\ref{radbr}). We note that \citet{1997ApJ...475..786G} considered $k$ values above 0.15 in the case of a Wolf-Rayet wind, due to the higher metallicity as compared to their O-type progenitors, while \citet{1982ApJ...259..282A} proposed values of 0.174 and 0.178 for O-stars with effective temperature of 40000 and 50000 K, respectively (more typical of the HD\,93205's primary). The latter values lead to a braking parameter greater than one in a significant part of the orbit.

Another radiative effect worth considering is {\it radiative inhibition} \citep{1994MNRAS.269..226S}. This process considers the mutual influence of the radiation pressure of both stars in a massive binary system on their wind material. Basically, the incoming primary wind is likely to be slower toward the direction of the secondary due to the continuous contribution of the secondary radiation field on the primary wind equation of motion. Such a process is known to be able to reduce substantially the pre-shock velocity of a colliding wind, therefore significantly reducing the post-shock temperature of the X-ray emitting plasma. In case of substantial inhibition, the effective wind velocity could drop by 50\,$\%$ (see examples in \citealt{1994MNRAS.269..226S}).

Finally, a likely radiative effect is that of {\it self-regulated shocks} \citep{2013ApJ...767..114P}. In this case, the X-ray emission produced by the post-shock plasma ionizes the incoming wind ahead of the shock, leading to significant inhibition of the wind acceleration. The effect is twofold: (i) it can reduce the pre-shock velocity, therefore reducing the post-shock plasma temperature, and (ii) the ionization close to the shock in the pre-shock flow can make the line-driving mechanism much less efficient, thus potentially suppressing sudden radiative breaking. As a result, this process has the potential to extend the range of orbital separations where a WPC occurs. Consequently, even cases with CAK parameters in favor of sudden radiative breaking around apastron (Fig.\,\ref{radbr}, dotted line) may actually lead to a WPC along the full orbit.\\

Summarizing, radiative effects play a key role in the nature and properties of the wind interaction. The detailed analysis of their combined action would require a full hydro-radiative modelling, based on robust stellar and wind parameters for both components of the system. However, this can't prevent us from exploring potential scenarios expected to be constrained by our X-ray measurements. According to our analysis, the WPC scenario is worth considering. However, depending on the adopted initial parameters, one cannot fully reject the idea of a WWC. This is especially true if one reduces the primary's wind velocity and/or mass loss rate, and alternative assumptions are adopted for the line-driving parameters affecting the occurrence of radiative braking. Finally, as a result of the significant eccentricity of the system, the pre-shock conditions are changing along the orbit. We thus also have to consider a hybrid collision scenario (HCP) where the system is switching between a WPC (around periastron) and a WWC scenario (around apastron). 

\subsection{Wind-photosphere collision}\label{wpc}
\subsubsection{X-ray emission from a WPC}

\begin{figure}
\centering
    \includegraphics[width=1.0\columnwidth]{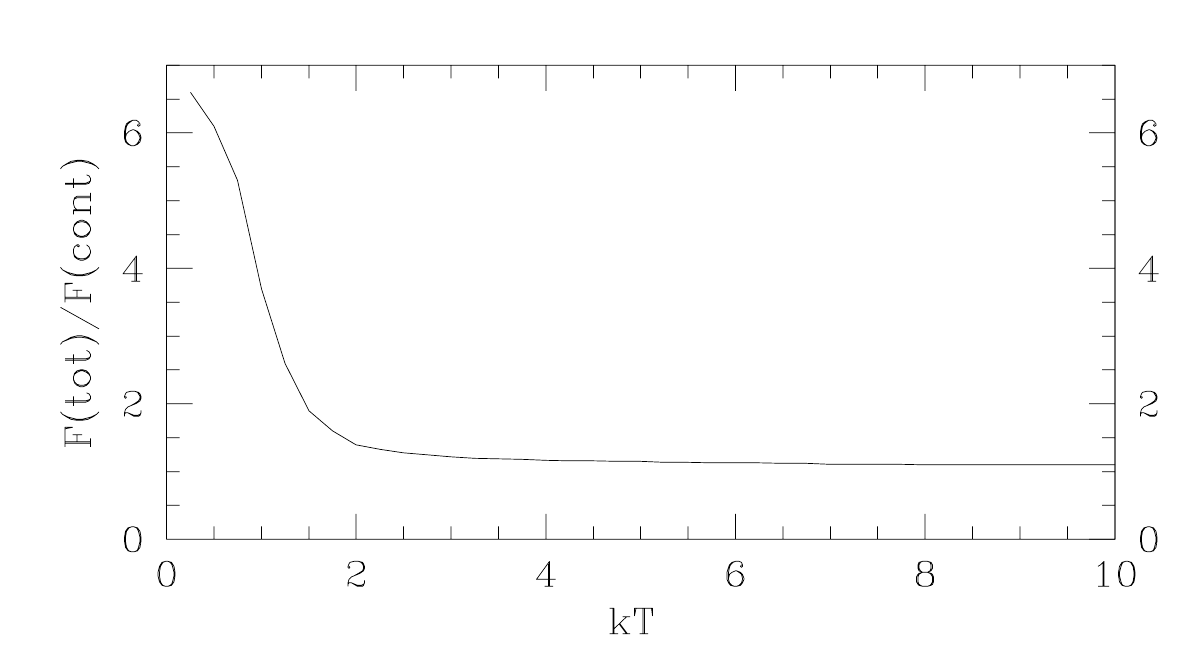}
	\caption{Plot of the total-to-continuum X-ray flux ratio as a function of plasma temperature.\label{fratio}}
\end{figure}

In the WPC case, a shock warped about the secondary star is formed. The pre-shock flow is made up by the incoming primary wind, and the heated post-shock gas is between the shock front and the secondary's surface. As a first step, we achieved some quantitative predictions of the X-ray luminosity from the WPC ($L_{X,WPC}$) based on the approach proposed by \citet[][their Eq.\,81]{1992ApJ...389..635U}. The predicted X-ray luminosity is expressed as follows:
\begin{equation}\label{wpcusov}
L_{X,WPC} = 5.8\,\times\,10^{32}\,\frac{{\dot M_{P,-5}}^2\,R_{S,12}^3}{D_{13}^4\,V_{P,8}}\,\,\,\,\,\,\,\mathrm{(erg\,s^{-1})}
\end{equation}
\noindent where ${\dot M_{P,-5}}$ is the primary wind mass loss rate in units of 10$^{-5}$\,M$_\odot$\,yr$^{-1}$, $R_{S,12}$ is the secondary's radius in units of 10$^{12}$\,cm, and $V_{P,8}$ is the pre-shock velocity of the primary's wind in units of 10$^{8}$\,cm\,s$^{-1}$. The separation $D_{13}$ (in units of 10$^{13}$\,cm) is determined as a function of the orbital phase on the basis of the parameters provided in Table\,\ref{hd93205_orb_par1}. 

We stress that Eq.\,\ref{wpcusov} established by \citet{1992ApJ...389..635U} considers the thermal X-ray emission produced by the post-shock plasma is continuum free-free emission only, and completely neglects the contribution from emission lines. Eq.\,\ref{wpcusov} thus underestimates the actual X-ray luminosity by a factor that depends on the post-shock plasma temperature, and thus on the pre-shock velocity, as expressed below \citep{1992ApJ...386..265S},
\begin{equation}\label{psT}
kT \approx 1.17\,V_{P,8}^2
\end{equation}
We quantified the relative importance of emission lines making use of thermal emission models in XSPEC. We measured the theoretical X-ray flux between 0.3 and 10.0 keV for a series of thermal emission models with consistent normalization parameter, for plasma temperatures ($kT$) ranging between 0.25 to 10.0 keV (by steps of 0.25 keV). We repeated the procedure using two models offered in XSPEC: (i) {\tt apec}, accounting for both free-free continuum and emission lines (total emission), and (ii) {\tt nlapec}, accounting free-free continuum with no contribution from emission lines. We computed the total-to-continuum flux ratio at each temperature, as shown in Fig.\,\ref{fratio}. The contribution from spectral lines clearly dominates at low plasma temperature, and drops quickly to reach a level of about 10\,$\%$ of the continuum contribution above 2\,keV. This correction has thus to be provided to $L_{X,WPC}$ depending on the post-shock temperature.

\begin{figure}
\centering
    \includegraphics[width=1.0\columnwidth]{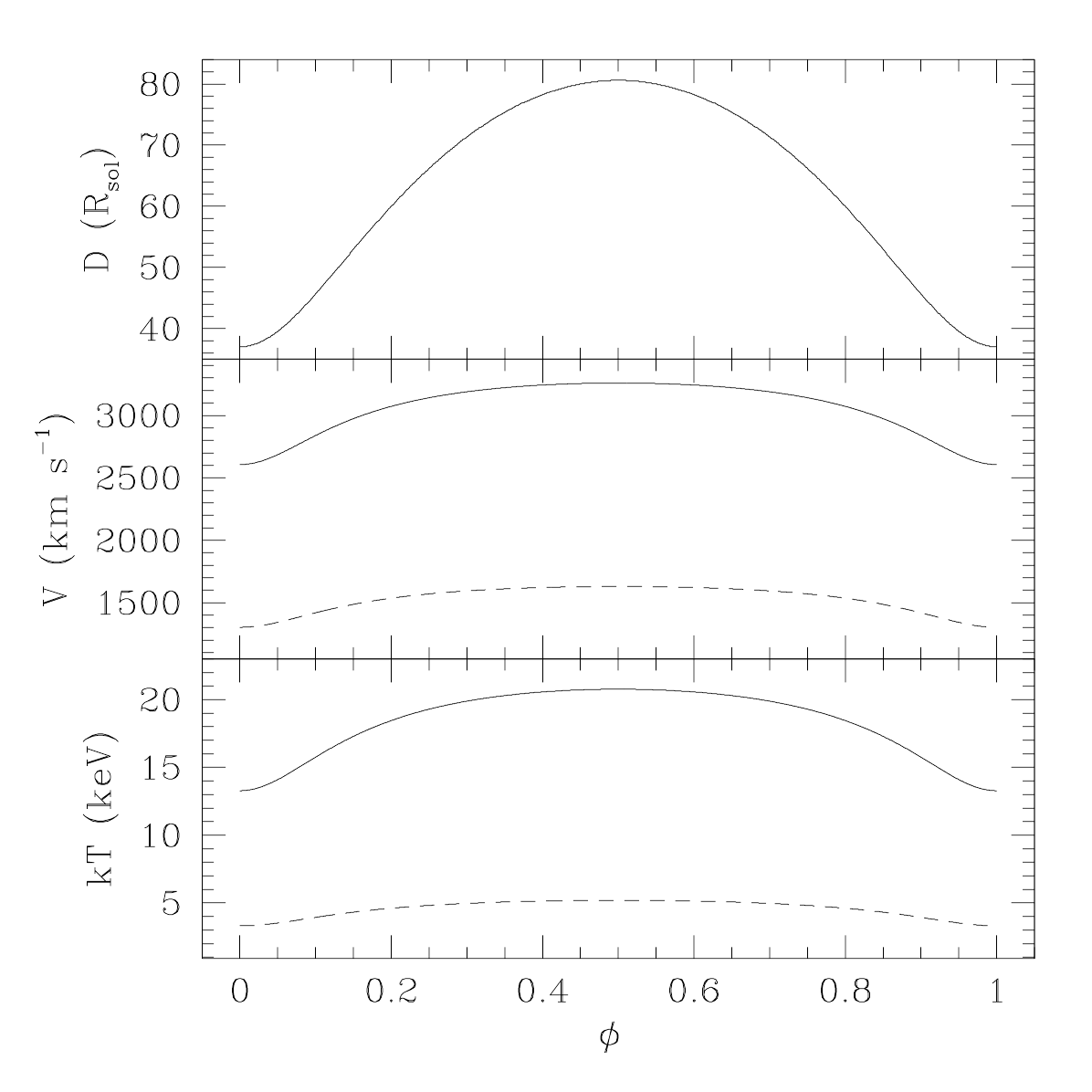}
	\caption{Plot of the stellar separation ({\it upper panel}), of the pre-shock velocity ({\it middle panel}) and post-shock temperature ({\it lower panel}) as a function of the orbital phase. Solid lines stand for a first approximation of the pre-shock velocity while dashed lines illustrate the case of strong radiative inhibition.\label{SVT}}
\end{figure}

The pre-shock velocity is varying significantly as a function of the orbital phase, as a consequence of the eccentricity of the short period orbit where the primary wind collides the photosphere while still in its acceleration region. Fig.\,\ref{SVT} shows the pre-shock velocity at the position of the secondary (middle panel, solid line), along with the post-shock plasma temperature (lower panel, solid line) as a function of the orbital phase. For illustration purpose, dashed lines for the velocity and temperature illustrate the hypothetical situation of a significant drop in the pre-shock velocity due to high amplitude radiative inhibition (see Sect.,\ref{radeffect}). 

\begin{figure}
\centering
    \includegraphics[width=1.0\columnwidth]{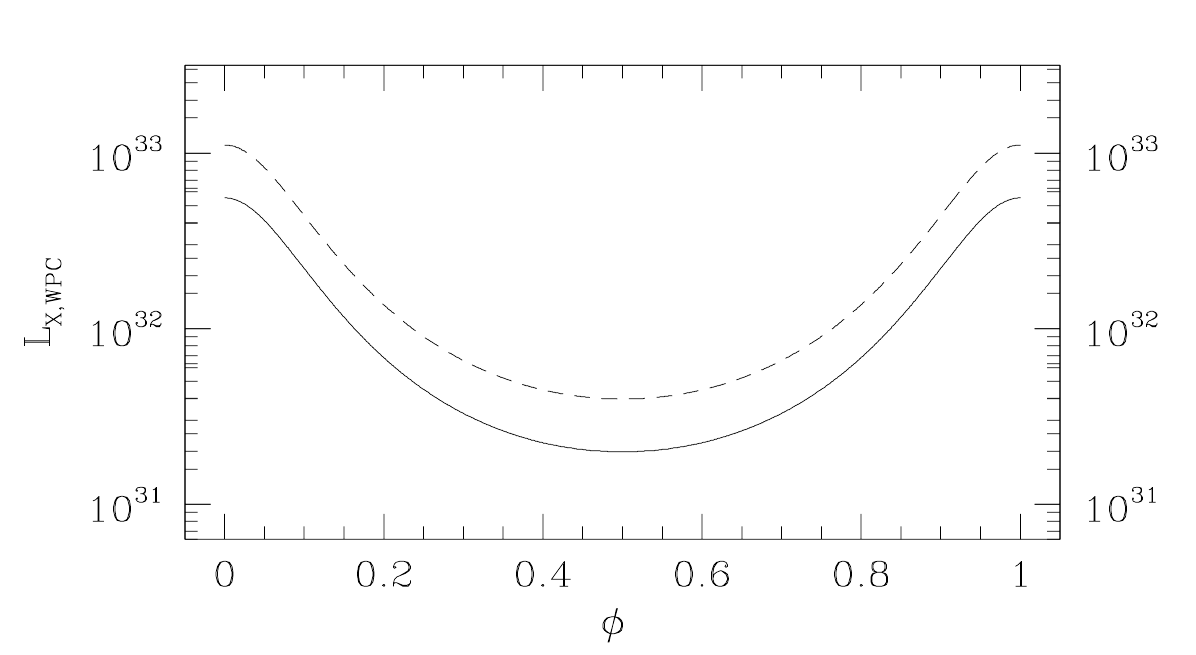}
	\caption{Plot of the predicted X-ray luminosity ($L_{X,WPC}$) as a function of the orbital phase. The solid and dashed line has the same meaning as in Fig.\,\ref{SVT}.\label{WPClum}}
\end{figure}

Based on the considerations on the pre-shock velocity, Eq.\,\ref{wpcusov} allows us to predict the X-ray luminosity from the WPC region, as shown in Fig.\,\ref{WPClum}. According to Fig.\,\ref{fratio}, the range of temperatures displayed in the bottom panel of Fig.\,\ref{SVT} suggests a quite hard X-ray emission, dominated by the free-free continuum with a contribution from spectral lines of some additional 10\,$\%$. We note however that these temperature are clearly quite excessive regarding the plasma temperature values we reported on in Sect.\,\ref{spec}. This point will be addressed in Sect.\,\ref{scenario}.

\subsubsection{Post-shock temperature}\label{shock}
According to our discussion in Sect.\,\ref{excess}, HD\,93205 may give rise to a WPC in a significant part of its orbit, or even along its full orbit. Focusing on this scenario, we explored the question of the likely shape of the shock front to investigate in particular the impact of shock obliquity on the distribution of the pre-shock velocities (and resulting post-shock temperatures) along the shock surface. Our methodology is presented in Appendix\,\ref{shapewpc}.

\begin{figure*}
\centering
\subfigure[Periastron]
{\includegraphics[width=1.0\columnwidth]{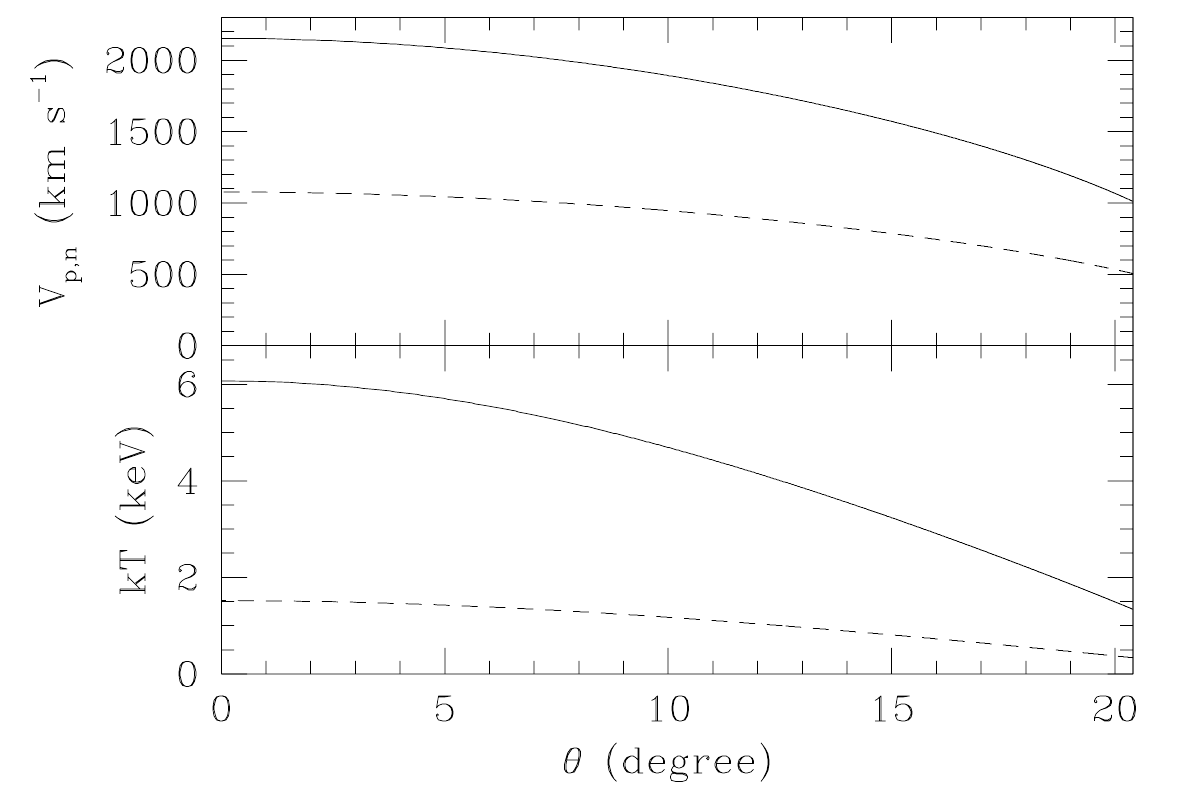}}
\subfigure[Apastron]
{\includegraphics[width=1.0\columnwidth]{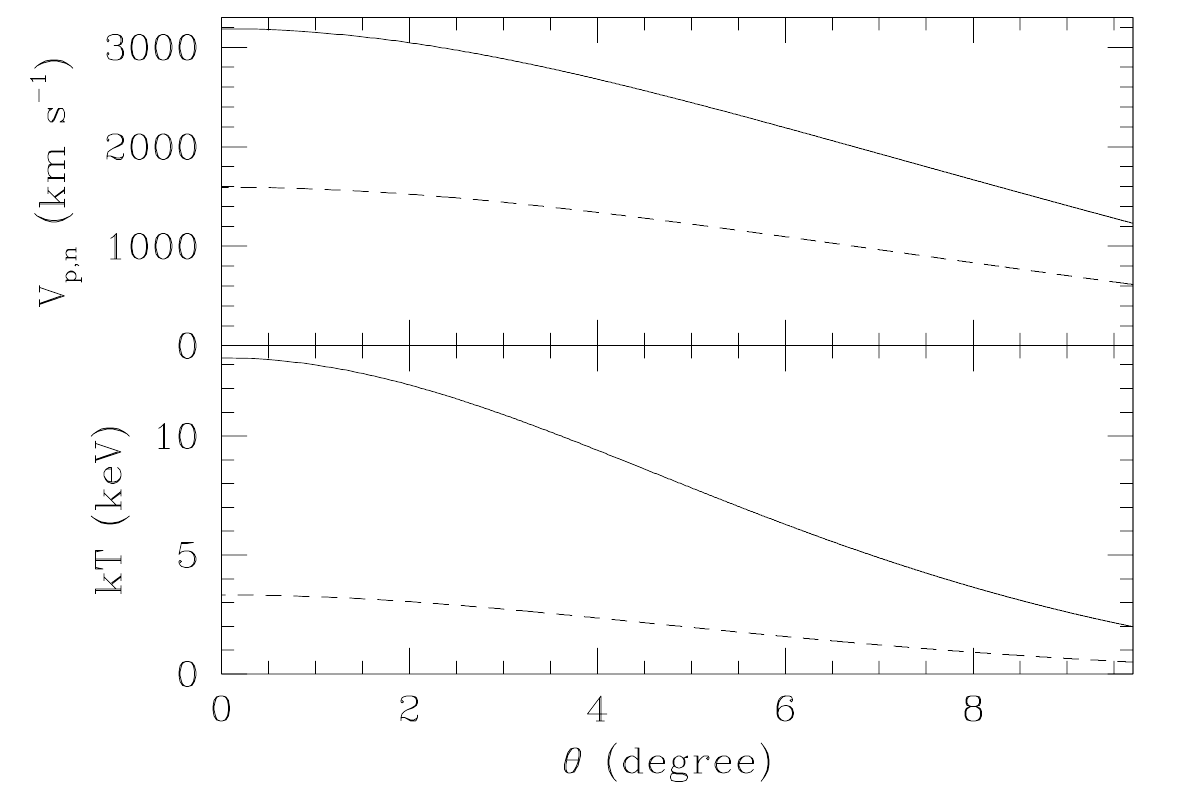}}
	\caption{Plot of the expected normal pre-shock velocity as a function of angular separation from the line of centers ($\theta$) at periastron ({\it left}) and apastron ({\it right}), in the framework of WPC scenario. Solid and dashed lines stand for the non-inhibited and strongly radiatively inhibited cases, respectively. \label{vshock}}
\end{figure*}

On the basis of normal pre-shock velocities determined as a function of $\theta$ (Eqs.\,\ref{phi} to \ref{theta}), we determined the expected post-shock temperature along the shock surface using Eq.\,\ref{psT}. We repeated the calculation considering both periastron and apastron, in both un-inhibited and strongly radiatively inhibited (wind velocity dropped by a factor 2) cases. We restricted the calculation up to off-axis angle ($\theta$) values corresponding to $x = 0$ (see Fig.\,\ref{bowshock}). This explains why the range of $\theta$ values displayed in Fig.\,\ref{vshock} is different for the two considered orbital phases (the shock is seen under a lower off-axis angle at apastron as compared to periastron). Depending on the orbital phase, on the considered $\theta$ value, and on the level of radiative inhibition, the range of normal pre-shock velocities is quite high, i.e. from about 500 to 3200 km\,s$^{-1}$. This leads to an even wider range of post-shock temperatures, from about 0.5\,keV to 14\,keV.

Because of the strong effect of obliquity along the shock front, the resulting average post-shock temperature is significantly lower than the one usually measured at the apex of the shock (along the line of centers). In an attempt to achieve a view of the averaged post-shock temperature more typical of the full emitting shock, we computed the surface of revolution of the shock curve (Eq.\,\ref{eqshock}) between $x = 0$ and $x = R_S + \Delta$, yielding a value of $A_{shock}$ = 830\,R$_\odot^2$ (by steps of $\delta\,x$ = 0.01\,$R_\odot$). The surface averaged $kT$ was computed using this relation,
\begin{equation}
<kT> = \frac{\sum_{i}A_i\,(kT)_i}{\sum_{i}A_i} = \frac{1}{A_{shock}}\,\sum_{i}A_i\,(kT)_i
\end{equation}
\noindent where $A_i$ is the annular element of surface at position $x_i$. As each surface element increases proportionally to $y$ as one moves from the shock apex to more off-axis contributions, the relative weight of cooler plasma is high enough to drop the average temperature significantly below that post-shock gas close to the line of centers. As a result, $<kT>$ goes from 3.0 to 8.0\,keV in the non-inhibited case at periastron and apastron, respectively. Assuming severe radiative inhibition (see above), these values drop down to 0.8 and 2.0 keV. In addition, our basic approach considers only the emission for the positive side of the $x$-axis. On the side of the back hemisphere of the secondary, some emission is also expected with a high obliquity, i.e. with a lower plasma temperature. However, we stress that for negative values of $x$, the primary wind flow line gets quickly close to the asymptote of the shock surface, leading to a simple advection of pre-shock material along the shock without feeding any further the shock physics at the origin of the X-ray emission. This asymptotic flow case is especially quickly reached for shorter separations. We thus anticipate that the X-ray contributions that are not explicitly considered above should mainly lead to a slight drop of the average plasma temperature away from periastron. In addition, away from the line of centers, the influence of the secondary wind can be sufficiently strong to contribute to the shaping of the shock front. Dealing with these aspects is however beyond the scope of this paper, as this would require a full hydrodynamical and radiative modelling of the system.

Finally, the likely wind velocity parameter space can encompass lower values if for instance the assumed terminal velocity is too high. In addition, the assumed $\beta$ parameter of the velocity law (Eq.\,\ref{betalaw}) has some influence on the results. For instance, for $\beta = 1.0$, $<kT>$ goes down to 2.4 (resp. 7.6) at periastron (resp. apastron) for the un-inhibited case. For severe inhibition (see above), these values drop by a factor 4. 

\subsection{Wind-wind collision}\label{wwc}
The case of WWC has been extensively discussed by seminal papers in this field \citep[see e.g.][]{1992ApJ...386..265S,2002A&A...388L..20P,2010MNRAS.403.1657P}. The occurrence of WWC is expected to significantly affect the X-ray properties of the system as compared to the WPC scenario. The main reason lies in the fact that such a WWC would occur in the acceleration region of the primary wind. As such a WWC would occur further away from the secondary as compared to the shock discussed in Sect.\,\ref{wpc}, the primary wind pre-shock velocity would be lower, leading to significantly lower post-shock temperatures. This is illustrated in Fig.\,\ref{wwc} where the velocity curves are plotted for both winds. Besides, a WWC would harbor two shocks, including one on the secondary's side. Given the proximity with the secondary star, this secondary shock would be characterized by low pre-shock velocities, resulting in a quite soft X-ray emission component from the post-shock secondary plasma. 

\begin{figure*}
\centering
\subfigure[Periastron]
{\includegraphics[width=1.0\columnwidth]{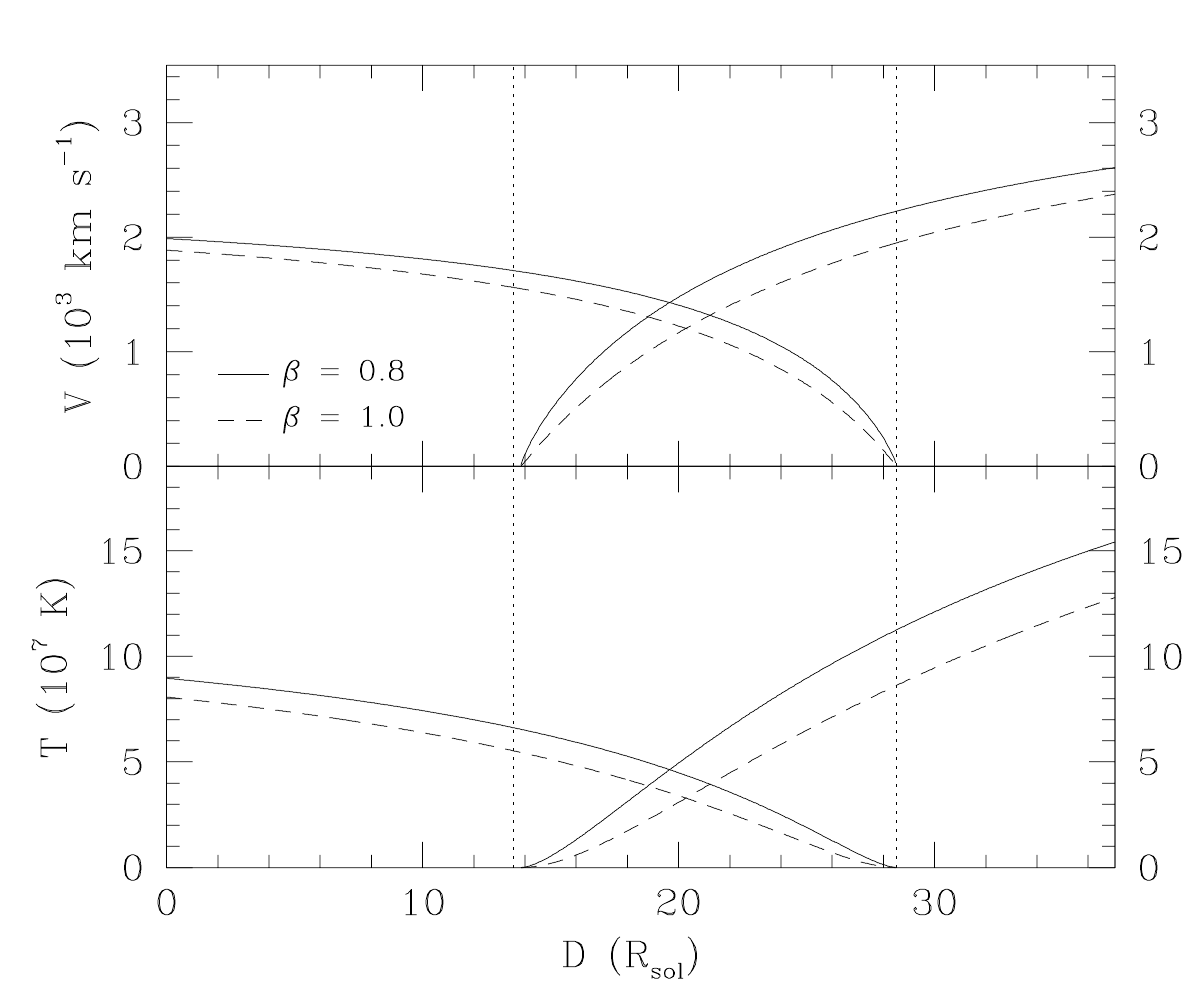}}
\subfigure[Apastron]
{\includegraphics[width=1.0\columnwidth]{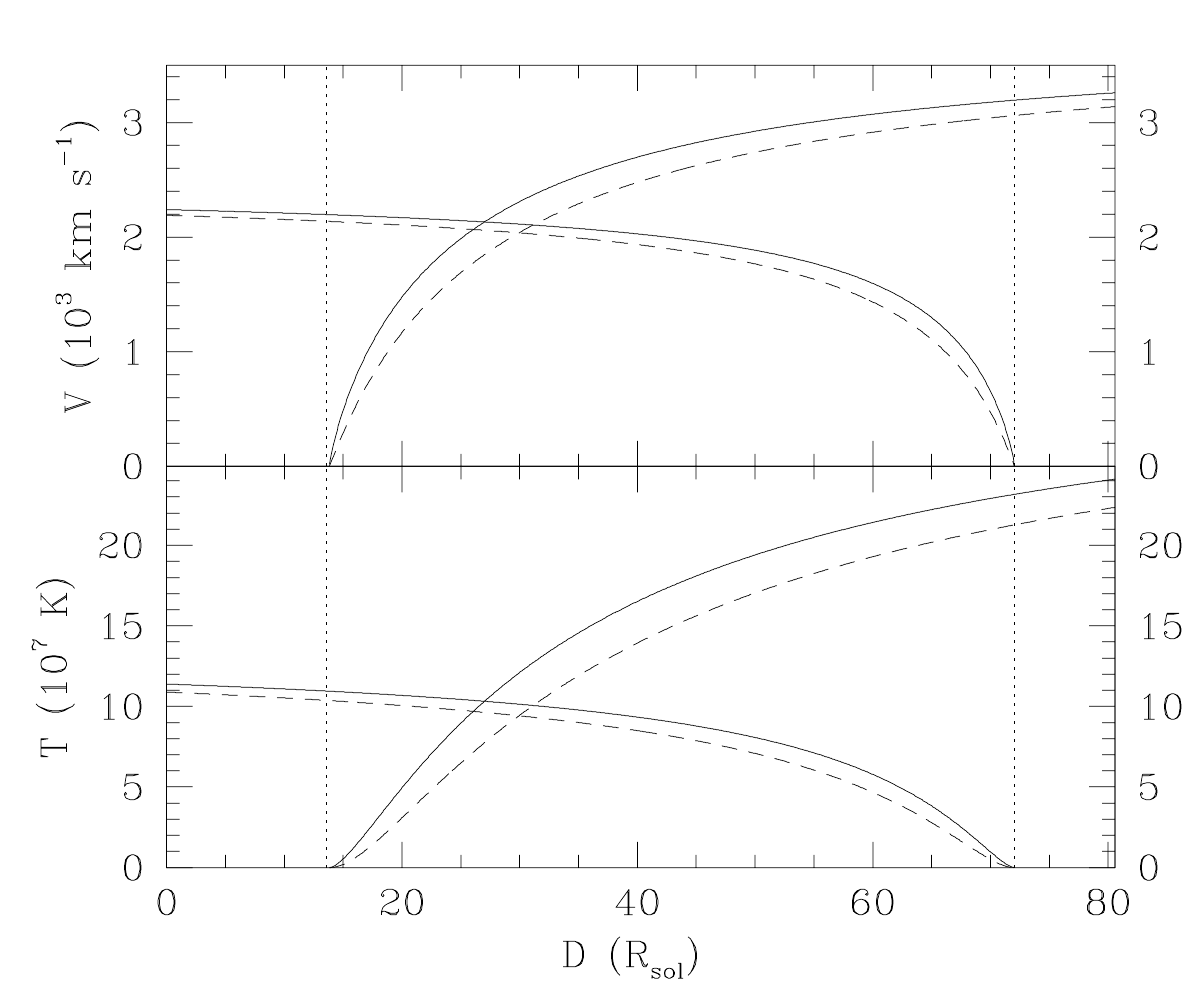}}
	\caption{{\it Top panels:} Velocity curves assuming two different values of the $\beta$ parameter for both winds: 0.8 (solid curves) and 1.0 (dashed curves). {\it Bottom panels:} Post-shock temperatures. The primary star is set at $D = 0$, and the secondary is at the other end of the horizontal axis. The vertical dotted lines delimit the maximum region available for wind acceleration between the stellar bodies along the line of centers.\label{wwc}}
\end{figure*}

Depending on the position of the WWC between the two stars, different ranges of pre-shock velocities could be covered. It is also noticeable the the small size of the orbit and its quite large eccentricity lead the distance between photospheres to range between $\sim$\,14.7\,R$_\odot$ at periastron and $\sim$\,58.2\,R$_\odot$ at apastron (i.e. a factor $\sim\,4$). This results in substantial differences in pre-shock velocities as a function of the orbital phase. If strong radiative inhibition was considered, curves in the upper panels of Fig.\,\ref{wwc} would be significantly dropped, resulting in a severe drop of the post-shock temperature curves (lower panels). Additionally, the velocity curves plotted in Fig.\,\ref{wwc} are valid along the line of centers. Accounting for obliquity due to the WWC warped about the secondary, one may expect the average X-ray emitting plasma temperature on the primary side to be dropped by about a factor 2. The obliquity is however expected to be less severe on the secondary's side, resulting in an obliquity reduction factor still lower than 1, but significantly larger than 0.5. This would overall constitute a way to predict plasma temperature closer to those measured from the X-ray emission in Sect.\,\ref{spec}.

\subsection{Hybrid wind collision}\label{hwc}
Depending on adopted values for stellar wind parameters and on the action of efficient radiative effects, a stable stagnation point between the two wind may exist at some orbital phases, especially away from periastron. Fig.\,\ref{radbr} (along with the discussion in Sect.\,\ref{radeffect}) also tells us that radiative braking may prevent the existence of WPC around apastron, over a part of the orbit that depends sensitively on adopted parameters. This scenario would therefore consider that the nature of the wind interaction would change over time scales shorter than the orbit. It is however very difficult to predict how such a wind collision system would relax from such a change of configuration. One can just anticipate a rather complex situation whose accurate description is far beyond the scope of this study. On the one hand, this would require a detailed time-dependent modelling of the hydrodynamic of the wind interaction, and on the other hand our current data are not sufficient to provide any insight into this topic.

\section{A potential scenario for HD\,93205}\label{scenario}

\subsection{Orbital modulation}

According to the results reported in Sect.\,\ref{lc}, the X-ray emission presents a clear orbital modulation. The maximum-to-minimum $L_{X,excess}$ for the broad band (see Sect.\,\ref{sectxlum}) is about 1.6. This means that the X-ray emission from the wind interaction region drops by a factor slightly greater than 1/3 from maximum to minimum. This phase-locked X-ray variability is in principle attributable to a variation of the emission (due to the significant eccentricity) modulated by orientation effects. 

\begin{figure*}
\centering
\subfigure[Orbit]
{\includegraphics[width=1.0\columnwidth]{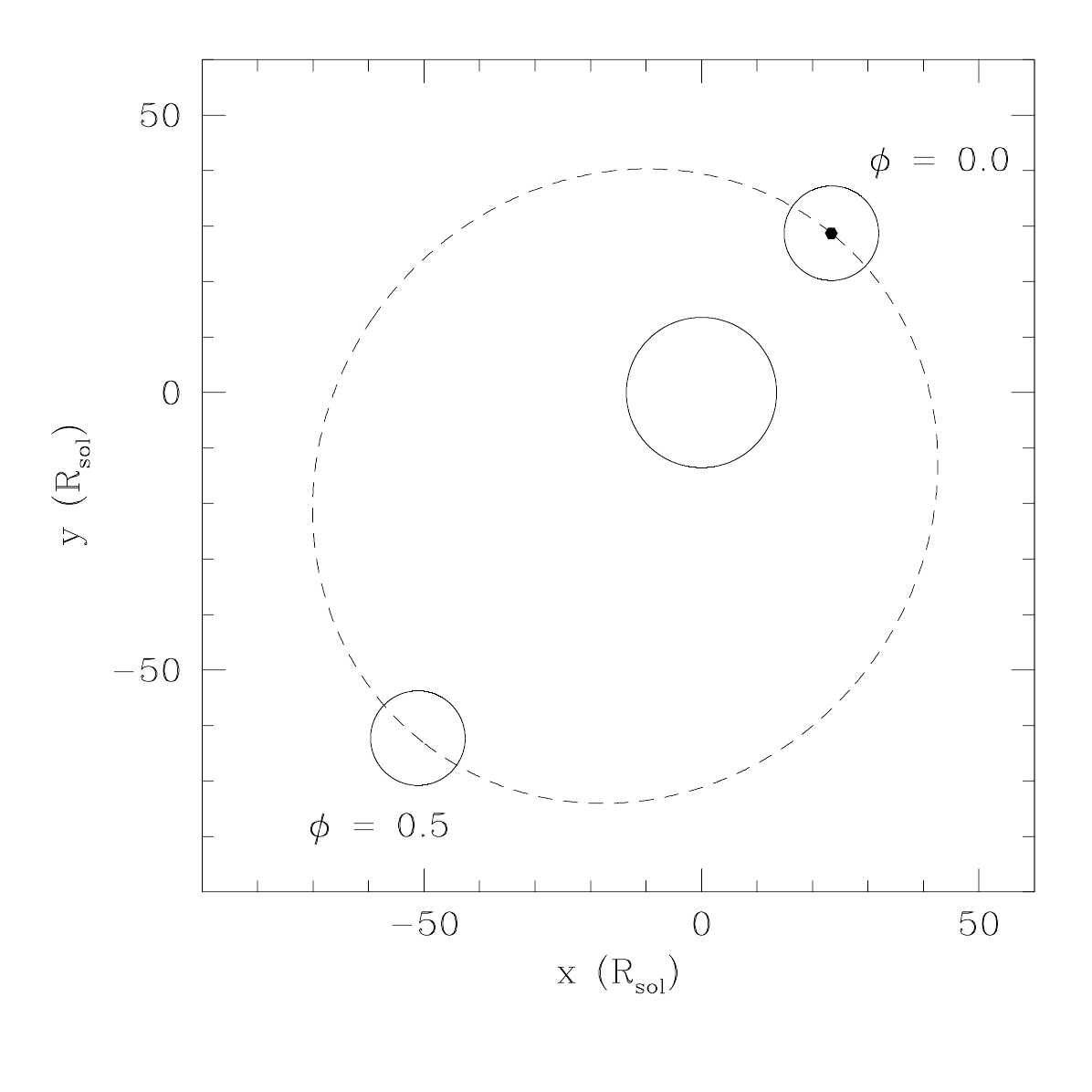}}
\subfigure[Projected orbit]
{\includegraphics[width=1.0\columnwidth]{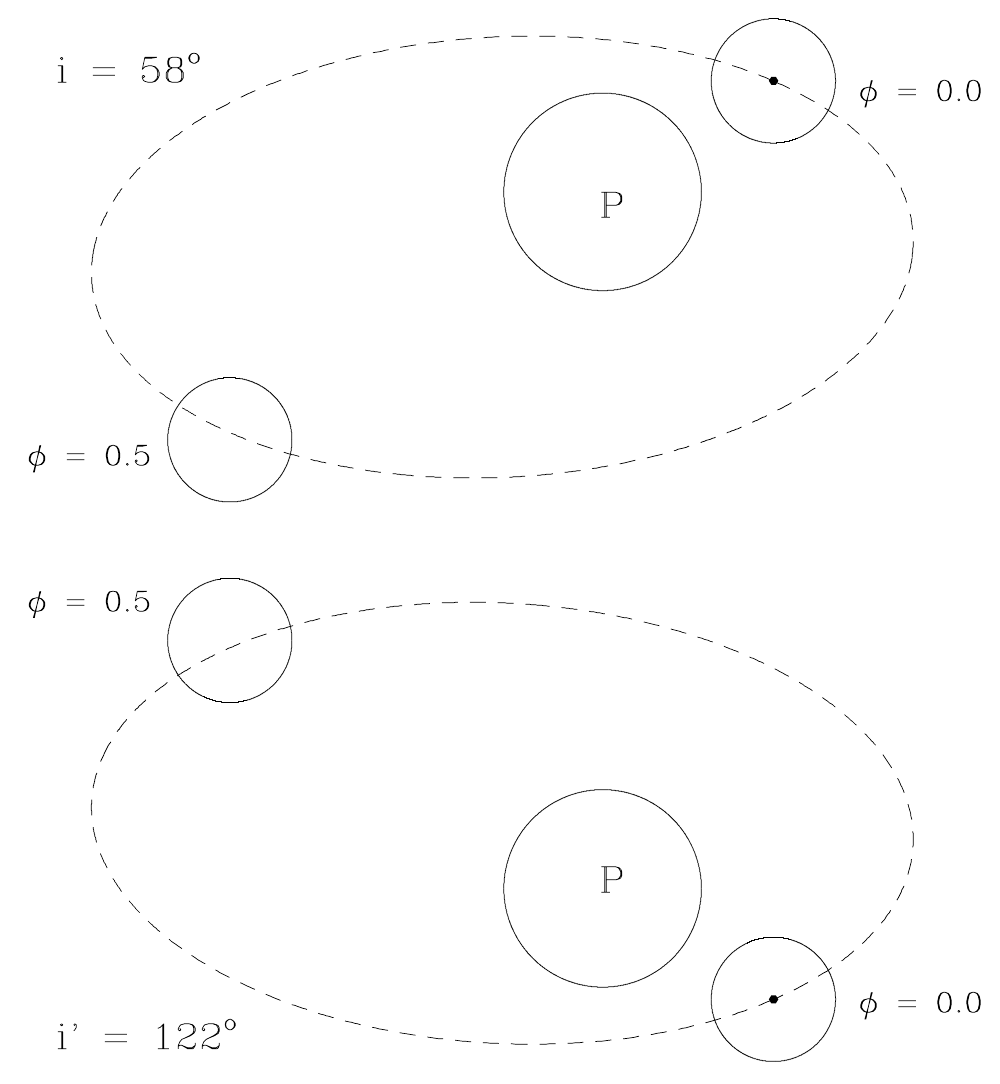}}
	\caption{Illustration of the orbit (dashed line) of HD\,93205 in the orbital plane with the primary set at coordinates ($x = 0,y = 0$) ({\it left}) and as a projected orbit assuming two likely inclination angles ({\it right}). Solid circles represent the primary and secondary stellar bodies. Stellar bodies and orbit sizes are to scale. Orbital phase increases counterclockwise (resp. clockwise) along the orbit in top right (resp. bottom right) illustration. \label{orbit}}
\end{figure*}

The intrinsic X-ray emission from the wind interaction region is expected to peak close to periastron, when the emission measure of the emitting plasma is at maximum. In the framework of the WPC scenario, the expected minimum occurs indeed at apastron (Fig.\,\ref{WPClum}). Assuming a WWC scenario, one would expect a similar behaviour \citep[see e.g.][]{2010MNRAS.403.1657P}. Even though our light curves display a peak close to periastron, we also notice a low-state spanning orbital phase between about 0.5 and 1.0, with a steep rise at periastron followed by a shallower decrease of the X-ray flux. This asymmetric behaviour certainly points to a modulation by orientation effects.

The discussion of orientation effects in HD\,93205 requires having a close look at the orbit. Using the orbital elements published by \citet{2001MNRAS.326...85M}, we plotted the orbit of the system in Fig.\,\ref{orbit}. As the longitude of the ascending node is not known for HD\,93205, we set it to $0^\circ$. This however doesn't affect our discussion as this parameter only rules the rotation of the projected orbit on the sky plane. The left part of the figure shows the system in the orbital plane. The right part illustrates the projected orbit making two different assumptions on the actual inclination, as the available information doesn't allow to lift the ambiguity between $i = 58^\circ$ and $i' = 180^\circ - i = 122^\circ$. For both assumed inclination angles, the secondary is in the background at periastron and in the foreground at apastron.

One has to consider that the secondary stellar body is likely to hide a part of the X-ray emission region depending on the orbital phase, therefore leading to a drop of the X-ray flux. In addition, the X-ray emission is supposed to travel through an orbital phase dependent absorption column. The measured X-ray maximum occurs right after periastron, while the secondary's surface exposed to the primary wind is the most visible. The inclination prevents the secondary's surface to be eclipsed by the primary, even though one may expect some significant absorption by the primary wind. At minimum, after apastron, a significant part of the shocked region is occulted by the secondary's body. Given the results reported on here, it seems that the impact of the occultation is more severe than that of the absorption by the primary wind material. This suggests that, even though one could expect the X-ray emitting region to be somewhat extended, a significant part of the X-ray emitting wind interaction is confined in a narrow region close to the secondary's surface.

\subsection{Shock types}

\begin{figure}
\centering
    \includegraphics[width=1.0\columnwidth]{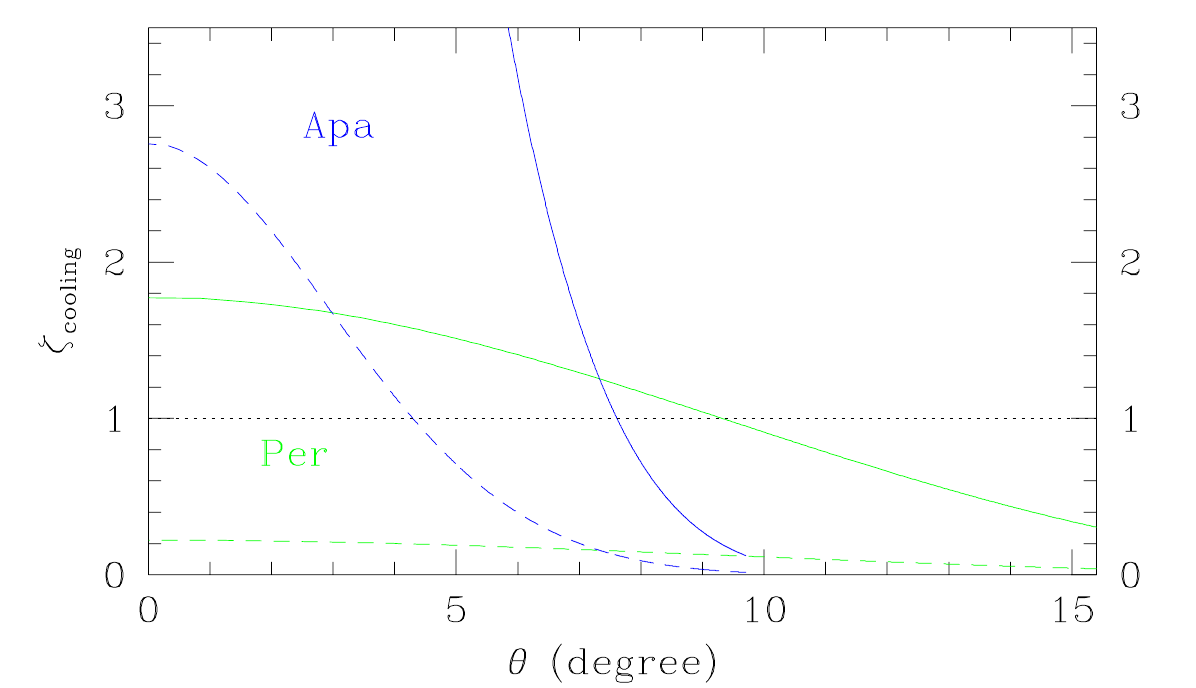}
	\caption{Plot of the cooling parameter ($\zeta_{cooling}$) as a function of the off-axis angle ($\theta$) for periastron (green) and apastron (blue), assuming the WPC scenario. The solid and dashed lines have the same meaning as in Fig.\,\ref{SVT}.\label{cooling}}
\end{figure}

Whatever the nature of the interaction (WPC or WWC), the nature of the shock (radiative vs adiabiatic) is also important to understand the X-ray emission from HD\,93205. We evaluated the cooling parameter ($\zeta_{cooling}$) as defined by \citet{1992ApJ...386..265S}. Assuming the WPC scenario, we calculated
\begin{equation}
\zeta_{cooling} = \frac{D_{12}\,V_{8}^4}{{\dot M}_{-7}}    
\end{equation}
\noindent where $D_{12}$ is the separation from the primary to a specific point of the shock front (P, in Fig.\,\ref{bowshock}) in units of 10$^{12}$\,cm, $V_{8}$ is the pre-shock velocity normal to the front at the same point in units of 10$^{8}$\,cm\,s$^{-1}$, and ${\dot M}_{-7}$ is the mass loss rate of the primary in units of $10^{-7}$\,M$_{\odot}$\,yr$^{-1}$. A $\zeta_{cooling}$ significantly above 1.0 points to an adiabatic regime, while a value close to or lower than 1.0 it typical of a radiative regime. Figure\,\ref{cooling} shows $\zeta_{cooling}$ as a function of the off-axis angles ($\theta$). At apastron, $\zeta_{cooling}$ is greater than 1.0 for $\theta$ values lower than 7.5$^\circ$ (resp. 4.0$^\circ$) in the non-inhibited (resp. severely radiatively inhibited) case. Sufficiently away from the line of centers, shock obliquity reduces sufficiently the pre-shock velocity to significantly affect $\zeta_{cooling}$. At periastron, $\zeta_{cooling}$ is of the order of (or even lower than) 1.0 over the whole shock front, even in the un-inhibited case.

The WWC scenario involves shocks occurring a bit further away from the secondary's photosphere. The pre-shock velocity on the primary's side is thus lower than in the case of the WPC scenario, and on the secondary's side a much lower pre-shock velocity is expected given the short available distance for wind acceleration (see Fig.\,\ref{wwc}). For both shocks, $\zeta_{cooling}$ is thus very low, given its strong dependence on the velocity. It is thus very reasonable to consider that both shocks will be radiative at any orbital phase, and for most (if not all) off-axis angles. The only exception could happen close to apastron, near the line of centers.

Finally, in the framework of HWC scenario, radiative braking could at best potentially prevent a photospheric collision to happen near apastron. Apastron would thus behave as a WWC system (where shocks are more likely radiative), and periastron would be more compliant with the WPC scenario (with radiative shocks). The main result is that the HWC scenario would very likely result in a fully radiative regime along the full orbit, whatever the position on the shock front.

The prevalence of the radiative nature of the shocks has inevitably consequences on the X-ray emission. The occurrence of the thin-shell instability is expected to alter the shock front, leading to an irregularly corrugated structure \citep{Vishniac1994,WF1998}. The resulting obliquity of the shocks causes a drastic drop of the pre-shock velocity, leading to significantly lower post-shock temperatures \citep{Kee2014}. This could explain the quite low post-shock temperature, i.e. about 0.2 and 0.6\,keV, measured across the orbit. On top of the likely radiative inhibition (see Sect.\,\ref{radeffect}), this additional obliquity should severely affect the energy injection of the primary wind into the shock front, therefore reducing substantially the heating of the X-ray emitting plasma. 

Coming again to the WPC scenario that should be valid along at least a major part of the orbit, Fig.\,\ref{WPClum} shows a predicted X-ray luminosity ranging between 6\,$\times$\,10$^{31}$ (at apastron) and 10$^{33}$\,erg\,s$^{-1}$ (at periastron), in the radiatively inhibited case. If we account for the significant drop of the pre-shock velocity due to the thin-shell instability, one has to consider a much lower average post-shock plasma temperature. Typically, at the maximum plasma temperature we measured in Sect.\,\ref{spec}, i.e. $\sim$ 0.6 keV, we have to multiply that predicted X-ray luminosity by a factor of about 5 to account for significant line emission on top of the continuum (see Fig.\,\ref{fratio} and discussion in Sect.\,\ref{wpc}). This leads to an X-ray luminosity ranging between 3\,$\times$\,10$^{32}$ (at apastron) and 5\,$\times$\,10$^{33}$\,erg\,s$^{-1}$ (at periastron). The former value is compliant with our measurements (Sect.\,\ref{sectxlum}), while the latter appears to be quite excessive. The lack of significant rise of the X-ray luminosity at peristron could be attributable to a significant photoelectic absorption by the primary wind, but it may also result from the radiative nature of the shocks, more pronounced close to periastron. According to 2D simulations of collision between equal expanding winds by \citet{Kee2014}, such radiative shocks should be less efficient at producing X-rays, with a reduction of the X-ray luminosity by a factor up to a few tens in the most severe cases  \citep[see e.g.][]{2004A&A...416..221D,2004ApJ...611..434A}. The suppression of thermal X-rays in radiative shocks was also addressed by \citet{2018MNRAS.479..687S}. More recent hydrodynamic simulations of \citet{2020MNRAS.492.5261K} for colliding wind structure of the massive binary system HD\,166734 (O7.5If $+$ O9I(f), period $\simeq$ 34.538 d, eccentricity $\simeq$ 0.618) along its binary orbit have also suggested formation of instabilities in the non-adiabatic interaction of unequal winds. The resulting X-ray emission is seen to be reduced as a consequence of accretion of primary wind onto secondary close to the periastron.

%______________________________________________________________

\section{Summary and conclusions}\label{conc}
We investigated the X-ray emission from the eccentric massive binary HD\,93205 on the basis of a time series of {\it XMM-Newton} observations spanning about two decades. The X-ray spectrum is quite soft, with a typical plasma temperature not higher than 0.6\,keV. We identified a periodic variability of the X-ray flux with a period of about 6.08\,d, in full agreement with the orbital solution published by \citet{2001MNRAS.326...85M}. 

The stellar wind of the primary component (O3.5\,V((ff))) clearly dominates that of the secondary (O8\,V), inhibiting the formation of a regular wind-wind interaction in between the two stars. Taking into account radiative effect such as sudden radiative breaking, radiative inhibition and the potential role played by self-regulated shocks, we conclude that HD\,93205 very likely undergoes a wind-photosphere interaction along most of its orbit. We caution however that this statement is supported by the one-dimensional approach used to address this topic, and we cannot reject the idea of a potential revision of this conclusion upon a more detailed three-dimensional analysis that is out of the scope of the present paper. The only exception is close to apastron, where one cannot exclude a wind-wind interaction to occur. One could thus be dealing with a hybrid wind collision, switching between a wind-photosphere collision and a wind-wind collision depending on the orbital phase. The orbital modulation of the X-ray emission is interpreted in terms variation of the intrinsic emission modulated by orientation effects (photoelectric absorption and occultation of a part of the X-ray emitting region by secondary stellar body). The X-ray light curve peaks close to periastron, with a steep rise at periastron followed by a shallower decay. The low-state of the X-ray emission occurs between phases 0.5 and 1.0, while the secondary is in front. The latter configuration implies that a significant part of the X-ray emitting interaction region is located on the opposite of the secondary stellar body (close to the line of centers, between the two stars). This is in agreement with our prediction of an interaction region (WPC of WWC) very close to the secondary's surface.

Despite the short separation in the system allowing at least the primary wind to hit the secondary's photosphere before it reaches its terminal velocity, predicted pre-shock velocities are too high to be compliant with the very soft emission revealed by the X-ray monitoring. Even considering a quite severe radiative inhibition dropping the pre-shock velocity by a factor 2, the predicted X-ray emission is harder than measured. This discrepancy could be explained by the radiative nature of the shock. In such shocks, instabilities significantly alter the shape of the shock front, leading to some additional obliquity in the incidence of the wind flow onto the shock surface.

Our study clearly demonstrates that HD\,93205 is a quite challenging laboratory to investigate shock physics in short period, asymmetric binary systems, where various radiative effects and radiative shocks concur to display an interesting behaviour in soft X-rays. We encourage to consider this target as a test-case for state-of-the-art future modelling including self-consistently all the physics that we considered in our discussion.

\begin{acknowledgements}
We thank the referee for the careful reading of the manuscript and giving us constructive comments and suggestions. BA acknowledges the grant provided by Wallonia Brussels International to carry out this work. This research has made use of observations obtained with \textit{XMM$-$Newton}, an ESA science mission with instruments and contributions directly funded by ESA Member States and NASA. 
\end{acknowledgements}

% WARNING
%-------------------------------------------------------------------
% Please note that we have included the references to the file aa.dem in
% order to compile it, but we ask you to:
%
% - use BibTeX with the regular commands:
%   \bibliographystyle{aa} % style aa.bst
%   \bibliography{Yourfile} % your references Yourfile.bib
%
% - join the .bib files when you upload your source files
%-------------------------------------------------------------------

\bibliographystyle{aa}

%\bibliography{ref2}{}

\begin{thebibliography}{53}
\expandafter\ifx\csname natexlab\endcsname\relax\def\natexlab#1{#1}\fi

\bibitem[{{Abbott}(1982)}]{1982ApJ...259..282A}
{Abbott}, D.~C. 1982, \apj, 259, 282

\bibitem[{{Albacete-Colombo} {et~al.}(2008){Albacete-Colombo}, {Damiani},
  {Micela}, {Sciortino}, \& {Harnden}}]{2008A&A...490.1055A}
{Albacete-Colombo}, J.~F., {Damiani}, F., {Micela}, G., {Sciortino}, S., \&
  {Harnden}, F.~R., J. 2008, \aap, 490, 1055

\bibitem[{{Albacete Colombo} {et~al.}(2003){Albacete Colombo}, {M{\'e}ndez}, \&
  {Morrell}}]{2003MNRAS.346..704A}
{Albacete Colombo}, J.~F., {M{\'e}ndez}, M., \& {Morrell}, N.~I. 2003, \mnras,
  346, 704

\bibitem[{{Anders} \& {Grevesse}(1989)}]{1989GeCoA..53..197A}
{Anders}, E. \& {Grevesse}, N. 1989, \gca, 53, 197

\bibitem[{{Antokhin} {et~al.}(2004){Antokhin}, {Owocki}, \&
  {Brown}}]{2004ApJ...611..434A}
{Antokhin}, I.~I., {Owocki}, S.~P., \& {Brown}, J.~C. 2004, \apj, 611, 434

\bibitem[{{Antokhin} {et~al.}(2003){Antokhin}, {Rauw}, {Vreux}, \& {van der
  Hucht}}]{2003ASPC..305..383A}
{Antokhin}, I.~I., {Rauw}, G., {Vreux}, J.~M., \& {van der Hucht}, K.~A. 2003,
  in Astronomical Society of the Pacific Conference Series, Vol. 305, Magnetic
  Fields in O, B and A Stars: Origin and Connection to Pulsation, Rotation and
  Mass Loss, ed. L.~A. {Balona}, H.~F. {Henrichs}, \& R.~{Medupe}, 383

\bibitem[{{Antokhin} {et~al.}(2008){Antokhin}, {Rauw}, {Vreux}, {van der
  Hucht}, \& {Brown}}]{2008A&A...477..593A}
{Antokhin}, I.~I., {Rauw}, G., {Vreux}, J.~M., {van der Hucht}, K.~A., \&
  {Brown}, J.~C. 2008, \aap, 477, 593

\bibitem[{{Arora} {et~al.}(2024){Arora}, {De Becker}, \&
  {Pandey}}]{2024A&A...687A..34A}
{Arora}, B., {De Becker}, M., \& {Pandey}, J.~C. 2024, \aap, 687, A34

\bibitem[{{Arora} {et~al.}(2019){Arora}, {Pandey}, \& {De
  Becker}}]{2019MNRAS.487.2624A}
{Arora}, B., {Pandey}, J.~C., \& {De Becker}, M. 2019, \mnras, 487, 2624

\bibitem[{{Arora} {et~al.}(2021){Arora}, {Pandey}, {De Becker}, {Pandey},
  {Chakradhari}, {Sharma}, \& {Kumar}}]{2021AJ....162..257A}
{Arora}, B., {Pandey}, J.~C., {De Becker}, M., {et~al.} 2021, \aj, 162, 257

\bibitem[{{Bailer-Jones} {et~al.}(2021){Bailer-Jones}, {Rybizki}, {Fouesneau},
  {Demleitner}, \& {Andrae}}]{2021AJ....161..147B}
{Bailer-Jones}, C.~A.~L., {Rybizki}, J., {Fouesneau}, M., {Demleitner}, M., \&
  {Andrae}, R. 2021, \aj, 161, 147

\bibitem[{{Billig}(1967)}]{Billig1967}
{Billig}, F.~S. 1967, Journal of Spacecraft and Rockets, 4, 822

\bibitem[{{Broos} {et~al.}(2011){Broos}, {Townsley}, {Feigelson}, {Getman},
  {Garmire}, {Preibisch}, {Smith}, {Babler}, {Hodgkin}, {Indebetouw}, {Irwin},
  {King}, {Lewis}, {Majewski}, {McCaughrean}, {Meade}, \&
  {Zinnecker}}]{2011ApJS..194....2B}
{Broos}, P.~S., {Townsley}, L.~K., {Feigelson}, E.~D., {et~al.} 2011, \apjs,
  194, 2

\bibitem[{Cardona \& Lago(2023)}]{CARDONA2023129185}
Cardona, V. \& Lago, V. 2023, Physics Letters A, 491, 129185

\bibitem[{{Castor} {et~al.}(1975){Castor}, {Abbott}, \&
  {Klein}}]{1975ApJ...195..157C}
{Castor}, J.~I., {Abbott}, D.~C., \& {Klein}, R.~I. 1975, \apj, 195, 157

\bibitem[{{Chlebowski} \& {Garmany}(1991)}]{1991ApJ...368..241C}
{Chlebowski}, T. \& {Garmany}, C.~D. 1991, \apj, 368, 241

\bibitem[{{Conti} \& {Walborn}(1976)}]{1976ApJ...207..502C}
{Conti}, P.~S. \& {Walborn}, N.~R. 1976, \apj, 207, 502

\bibitem[{{Corcoran}(1996)}]{1996RMxAC...5...54C}
{Corcoran}, M.~F. 1996, in Revista Mexicana de Astronomia y Astrofisica
  Conference Series, Vol.~5, Revista Mexicana de Astronomia y Astrofisica
  Conference Series, ed. V.~{Niemela}, N.~{Morrell}, P.~{Pismis}, \&
  S.~{Torres-Peimbert}, 54--60

\bibitem[{{Corcoran} {et~al.}(1995){Corcoran}, {Swank}, {Rawley}, {Petre},
  {Schmitt}, \& {Day}}]{1995RMxAC...2...97C}
{Corcoran}, M.~F., {Swank}, J., {Rawley}, G., {et~al.} 1995, in Revista
  Mexicana de Astronomia y Astrofisica Conference Series, Vol.~2, Revista
  Mexicana de Astronomia y Astrofisica Conference Series, ed. V.~{Niemela},
  N.~{Morrell}, \& A.~{Feinstein}, 97

\bibitem[{{De Becker}(2013)}]{2013NewA...25....7D}
{De Becker}, M. 2013, \na, 25, 7

\bibitem[{{De Becker}(2015)}]{2015MNRAS.451.1070D}
{De Becker}, M. 2015, \mnras, 451, 1070

\bibitem[{{De Becker} {et~al.}(2004){De Becker}, {Rauw}, {Pittard}, {Antokhin},
  {Stevens}, {Gosset}, \& {Owocki}}]{2004A&A...416..221D}
{De Becker}, M., {Rauw}, G., {Pittard}, J.~M., {et~al.} 2004, \aap, 416, 221

\bibitem[{{Gagn{\'e}} {et~al.}(2011){Gagn{\'e}}, {Fehon}, {Savoy}, {Cohen},
  {Townsley}, {Broos}, {Povich}, {Corcoran}, {Walborn}, {Remage Evans},
  {Moffat}, {Naz{\'e}}, \& {Oskinova}}]{2011ApJS..194....5G}
{Gagn{\'e}}, M., {Fehon}, G., {Savoy}, M.~R., {et~al.} 2011, \apjs, 194, 5

\bibitem[{{Gayley} {et~al.}(1997){Gayley}, {Owocki}, \&
  {Cranmer}}]{1997ApJ...475..786G}
{Gayley}, K.~G., {Owocki}, S.~P., \& {Cranmer}, S.~R. 1997, \apj, 475, 786

\bibitem[{{Horne} \& {Baliunas}(1986)}]{1986ApJ...302..757H}
{Horne}, J.~H. \& {Baliunas}, S.~L. 1986, \apj, 302, 757

\bibitem[{{Jansen} {et~al.}(2001){Jansen}, {Lumb}, {Altieri}, {Clavel}, {Ehle},
  {Erd}, {Gabriel}, {Guainazzi}, {Gondoin}, {Much}, {Munoz}, {Santos},
  {Schartel}, {Texier}, \& {Vacanti}}]{2001A&A...365L...1J}
{Jansen}, F., {Lumb}, D., {Altieri}, B., {et~al.} 2001, \aap, 365, L1

\bibitem[{{Jenkins}(2019)}]{2019ApJ...872...55J}
{Jenkins}, E.~B. 2019, \apj, 872, 55

\bibitem[{{Kashi}(2020)}]{2020MNRAS.492.5261K}
{Kashi}, A. 2020, \mnras, 492, 5261

\bibitem[{{Kee} {et~al.}(2014){Kee}, {Owocki}, \& {ud-Doula}}]{Kee2014}
{Kee}, N.~D., {Owocki}, S., \& {ud-Doula}, A. 2014, \mnras, 438, 3557

\bibitem[{{Lomb}(1976)}]{1976Ap&SS..39..447L}
{Lomb}, N.~R. 1976, \apss, 39, 447

\bibitem[{{Martins} {et~al.}(2005){Martins}, {Schaerer}, \&
  {Hillier}}]{2005A&A...436.1049M}
{Martins}, F., {Schaerer}, D., \& {Hillier}, D.~J. 2005, \aap, 436, 1049

\bibitem[{{Morrell} {et~al.}(2001){Morrell}, {Barb{\'a}}, {Niemela}, {Corti},
  {Albacete Colombo}, {Rauw}, {Corcoran}, {Morel}, {Bertrand}, {Moffat}, \&
  {St-Louis}}]{2001MNRAS.326...85M}
{Morrell}, N.~I., {Barb{\'a}}, R.~H., {Niemela}, V.~S., {et~al.} 2001, \mnras,
  326, 85

\bibitem[{{Muijres} {et~al.}(2012){Muijres}, {Vink}, {de Koter}, {M{\"u}ller},
  \& {Langer}}]{2012A&A...537A..37M}
{Muijres}, L.~E., {Vink}, J.~S., {de Koter}, A., {M{\"u}ller}, P.~E., \&
  {Langer}, N. 2012, \aap, 537, A37

\bibitem[{{Owocki} {et~al.}(2013){Owocki}, {Sundqvist}, {Cohen}, \&
  {Gayley}}]{2013MNRAS.429.3379O}
{Owocki}, S.~P., {Sundqvist}, J.~O., {Cohen}, D.~H., \& {Gayley}, K.~G. 2013,
  \mnras, 429, 3379

\bibitem[{{Parkin} \& {Sim}(2013)}]{2013ApJ...767..114P}
{Parkin}, E.~R. \& {Sim}, S.~A. 2013, \apj, 767, 114

\bibitem[{{Pittard} \& {Parkin}(2010)}]{2010MNRAS.403.1657P}
{Pittard}, J.~M. \& {Parkin}, E.~R. 2010, \mnras, 403, 1657

\bibitem[{{Pittard} \& {Stevens}(1997)}]{1997MNRAS.292..298P}
{Pittard}, J.~M. \& {Stevens}, I.~R. 1997, \mnras, 292, 298

\bibitem[{{Pittard} \& {Stevens}(2002)}]{2002A&A...388L..20P}
{Pittard}, J.~M. \& {Stevens}, I.~R. 2002, \aap, 388, L20

\bibitem[{{Roberts} {et~al.}(1987){Roberts}, {Lehar}, \&
  {Dreher}}]{1987AJ.....93..968R}
{Roberts}, D.~H., {Lehar}, J., \& {Dreher}, J.~W. 1987, \aj, 93, 968

\bibitem[{{Ryspaeva} \& {Kholtygin}(2020)}]{2020RAA....20..108R}
{Ryspaeva}, E. \& {Kholtygin}, A. 2020, Research in Astronomy and Astrophysics,
  20, 108

\bibitem[{{Scargle}(1982)}]{1982ApJ...263..835S}
{Scargle}, J.~D. 1982, \apj, 263, 835

\bibitem[{{Seward} \& {Chlebowski}(1982)}]{1982ApJ...256..530S}
{Seward}, F.~D. \& {Chlebowski}, T. 1982, \apj, 256, 530

\bibitem[{{Seward} {et~al.}(1979){Seward}, {Forman}, {Giacconi}, {Griffiths},
  {Harnden}, {Jones}, \& {Pye}}]{1979ApJ...234L..55S}
{Seward}, F.~D., {Forman}, W.~R., {Giacconi}, R., {et~al.} 1979, \apj, 234, L55

\bibitem[{{Sota} {et~al.}(2014){Sota}, {Ma{\'\i}z Apell{\'a}niz}, {Morrell},
  {Barb{\'a}}, {Walborn}, {Gamen}, {Arias}, \& {Alfaro}}]{2014ApJS..211...10S}
{Sota}, A., {Ma{\'\i}z Apell{\'a}niz}, J., {Morrell}, N.~I., {et~al.} 2014,
  \apjs, 211, 10

\bibitem[{{Steinberg} \& {Metzger}(2018)}]{2018MNRAS.479..687S}
{Steinberg}, E. \& {Metzger}, B.~D. 2018, \mnras, 479, 687

\bibitem[{{Stevens} {et~al.}(1992){Stevens}, {Blondin}, \&
  {Pollock}}]{1992ApJ...386..265S}
{Stevens}, I.~R., {Blondin}, J.~M., \& {Pollock}, A.~M.~T. 1992, \apj, 386, 265

\bibitem[{{Stevens} \& {Pollock}(1994)}]{1994MNRAS.269..226S}
{Stevens}, I.~R. \& {Pollock}, A.~M.~T. 1994, \mnras, 269, 226

\bibitem[{{Usov}(1992)}]{1992ApJ...389..635U}
{Usov}, V.~V. 1992, \apj, 389, 635

\bibitem[{{Vink} {et~al.}(2001){Vink}, {de Koter}, \&
  {Lamers}}]{2001A&A...369..574V}
{Vink}, J.~S., {de Koter}, A., \& {Lamers}, H.~J.~G.~L.~M. 2001, \aap, 369, 574

\bibitem[{{Vishniac}(1994)}]{Vishniac1994}
{Vishniac}, E.~T. 1994, \apj, 428, 186

\bibitem[{{Walborn}(1971)}]{1971ApJ...167L..31W}
{Walborn}, N.~R. 1971, \apjl, 167, L31

\bibitem[{{Walborn}(1973)}]{1973ApJ...179..517W}
{Walborn}, N.~R. 1973, \apj, 179, 517

\bibitem[{{Walder} \& {Folini}(1998)}]{WF1998}
{Walder}, R. \& {Folini}, D. 1998, \aap, 330, L21

\end{thebibliography}

\begin{appendix}
\section{Shock shape in the framework of the WPC scenario} \label{shapewpc}
As a first approximation of the shock shape, we made use of the shock stand-off distance and shape equation for a shock produced by a flow hitting a sphere \citep{Billig1967,CARDONA2023129185}. 

In cartesian coordinates with the $x$-axis along the line of centers of the binary, the equation expressing $y$ as a function of $x$ is the following,
\begin{equation}\label{eqshock}
y = \frac{R_c}{\tan\zeta}\,\Bigg[\Bigg(\frac{x - R_S - \Delta}{R_c\,\cotan^2\zeta} + 1\Bigg)^2 - 1\Bigg]^{1/2}
\end{equation}
\noindent where $R_S$ is the secondary stellar radius, $R_c$ is the shock curvature radius, $\Delta$ is the stand-off distance (separation between the shock apex and the stellar surface along the line of centers), and $\zeta$ is defined as
\begin{equation}
\zeta = \arcsin\bigg(\frac{1}{M_a}\bigg)
\end{equation}
\noindent where $M_a$ is the sonic pre-shock Mach number of the primary wind. 

The expressions for $R_c$ and $\Delta$ are the following:
\begin{equation}
R_c = 1.143\,R_S\,\exp\Bigg(\frac{0.54}{(M_a - 1)^{1.2}}\Bigg)
\end{equation}
\begin{equation}
\Delta = 0.143\,R_S\,\exp\Bigg(\frac{3.2}{M_a^2}\Bigg)
\end{equation}
Assuming a wind temperature that is 0.5 times the effective temperature, the sound speed (expressed as $v_s = 0.1\,\sqrt{T_{w,P}}$, assuming a perfect gas with mean molecular weight of the order of 1.0) is of the order of 15\,km\,s$^{-1}$. Given the expected range of pre-shock velocities as a function of orbital phase, $M_a$ ranges between 170 and 220. In that hypersonic regime, the shape and position of the shock are quite insensitive to changes of pre-shock sonic Mach number. Along the full orbit, we obtain $R_c = 9.75$\,R$_\odot$ and $\Delta = 1.22$\,R$_\odot$. We note that these values are still valid in case of severe radiative inhibition leading to a drop by a factor 2 of the primary wind velocity (as considered in Fig.\,\ref{SVT} and Fig.\,\ref{WPClum}). 

\begin{figure*}
\centering
    \includegraphics[width=1.8\columnwidth]{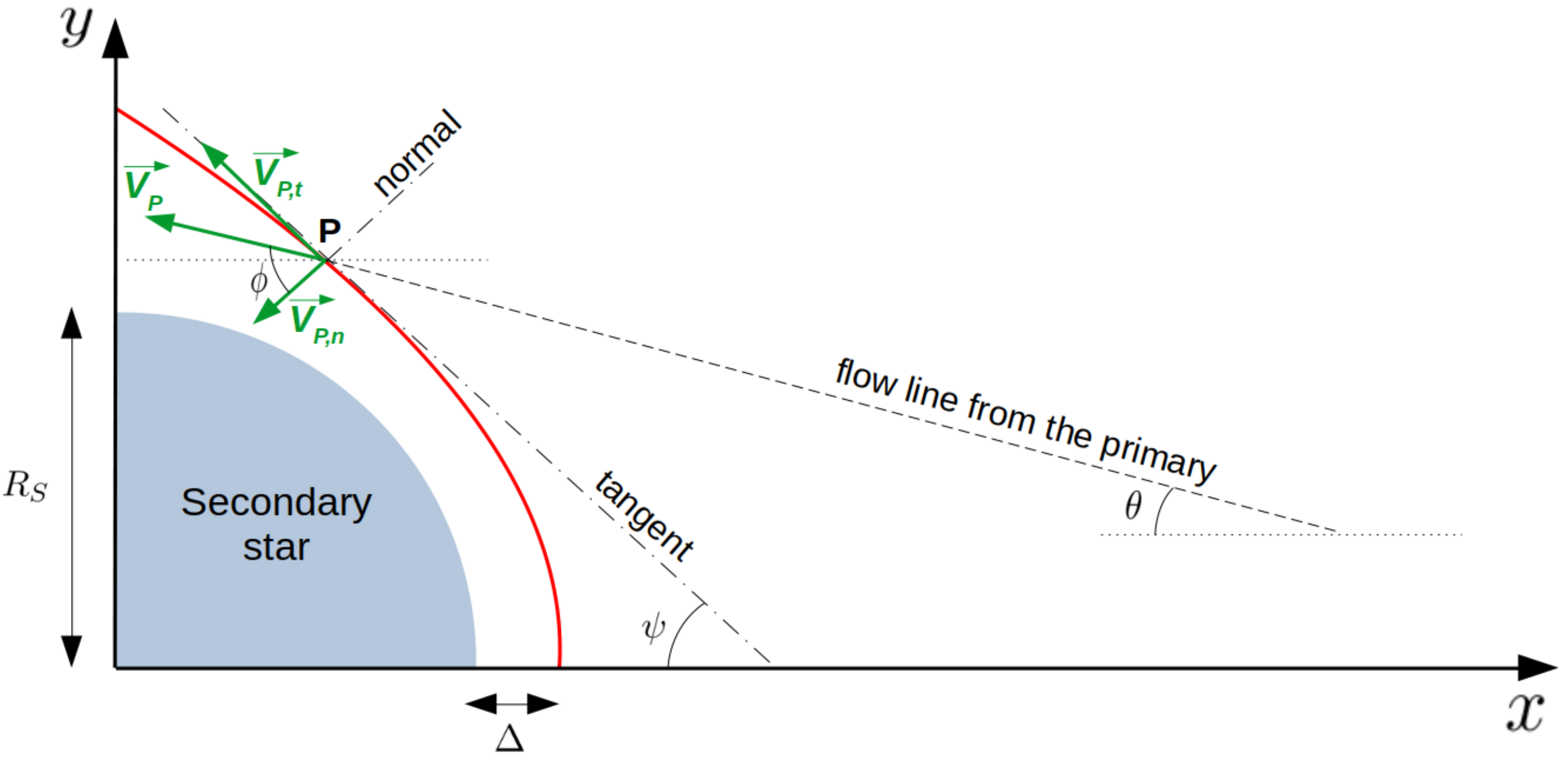}
	\caption{Illustration of the shock shape (red curve) produced by the WPC at an arbitrary orbital phase. Details required to compute the normal component of the pre-shock velocity are shown. The line of centers of the system is along the $x$-axis.\label{bowshock}}
\end{figure*}

An illustration of the shock shape is presented in Fig.\,\ref{bowshock}. The pre-shock velocity vector is represented for a specific flow line coming from the primary and hitting the shock at point P. The distance from the primary star to point P is easily expressed from the geometry of the system,
\begin{equation}
s = \sqrt{(D - X)^2 + y^2}
\end{equation}
Knowing $s$, the pre-shock velocity $V_P$ was computed using a standard $\beta$ velocity law (Eq.\,\ref{betalaw}).  

We made use of the geometry shown in Fig.\,\ref{bowshock} (based on Eq.\,\ref{eqshock}) to determine for every point of the shock curve the amplitude of the normal component of the pre-shock velocity ($V_{P,n}$) that counts for the shock physics at work in the system. With angles $\theta$ and $\psi$ defined as illustrated in Fig.\,\ref{bowshock}, we note that
\begin{equation}\label{phi}
\phi = \pi/2 - \psi + \theta 
\end{equation}
\noindent and
\begin{equation}\label{vpn}
V_{P,n} = V_P\,\cos\phi
\end{equation}
The angle $\theta$ is determined for any point P from the $x$ and $y$ coordinates related through Eq.\,\ref{eqshock},
\begin{equation}\label{theta}
\theta = \arctan\bigg(\frac{y}{D - x}\bigg)
\end{equation}

\begin{table*}
\caption{\label{tab:log} Log of \textit{XMM-Newton} observations of HD\,93205.}
\centering
\renewcommand{\arraystretch}{1.06}
\begin{tabular}{c c c c c c c c c}
\hline
\hline
 Sr. 	& Obs. ID & Detector  & Obs. Date  & Start time  & Duration\tablefootmark{b}   & Livetime\tablefootmark{c}   &Source\tablefootmark{d}     & Offset\tablefootmark{e}      \\
 No.    &         & (filter\tablefootmark{a})  &            & (UT)        &  (sec)    & (sec)      & counts    & (\arcmin)  	\\	
\hline
          1      & 0112580601    &     MOS1 (Th)       & 2000-07-26  & 04:59:20    &  36509  &  33060  &  1476$\pm$44   & 4.967 \\
                 &               &      PN  (Th)       &             &             &         &  27770  &  5834$\pm$89   &        \\
          2      & 0112580701    &     MOS1 (Th)       & 2000-07-27  & 23:49:25    &  12425  &  10900  &   400$\pm$23   & 4.967      \\
                 &               &      PN  (Th)       &             &             &         &   7912  &  1352$\pm$43   &        \\
          3      & 0112560101    &     MOS1 (Th)       & 2001-06-25  & 06:51:26    &  37052  &  28120  &  1751$\pm$47   & 3.132       \\ 
                 &               &     MOS2 (Th)       &             &             &         &  27870  &  1780$\pm$49   &        \\
                 &               &      PN  (Th)       &             &             &         &  20870  &  3917$\pm$69   &        \\
          4      & 0112560201    &     MOS1 (Th)       & 2001-06-28  & 07:22:56    &  40092  &  24120  &   990$\pm$37   & 3.132  \\  
                 &               &     MOS2 (Th)       &             &             &         &  24120  &  1054$\pm$41   &        \\
                 &               &      PN  (Th)       &             &             &         &  19850  &  2677$\pm$60   &        \\
          5      & 0112560301    &     MOS1 (Th)       & 2001-06-30  & 04:39:01    &  37714  &  36280  &  1466$\pm$45   & 3.132 \\
                 &               &     MOS2 (Th)       &             &             &         &  36320  &  1578$\pm$47   &        \\
                 &               &      PN  (Th)       &             &             &         &  30180  &  3965$\pm$72   &        \\            
          6      & 0145740101    &     MOS1 (Th)       & 2003-01-25  & 12:57:37    &  7252   &   6898  &   293$\pm$19   & 4.974  \\
                 &               &     MOS2 (Th)       &             &             &         &   6904  &   219$\pm$16   &        \\
          7      & 0145740201    &     MOS1 (Th)       & 2003-01-27  & 01:02:55    &  7250   &   6886  &   258$\pm$20   & 4.974  \\ 
                 &               &     MOS2 (Th)       &             &             &         &   6902  &   239$\pm$18   &        \\                   
          8      & 0145740301    &     MOS1 (Th)       & 2003-01-27  & 20:36:23    &  7247   &   6839  &   272$\pm$20   & 4.974 \\
                 &               &     MOS2 (Th)       &             &             &         &   6837  &   269$\pm$19   &        \\
          9      & 0145740401    &     MOS1 (Th)       & 2003-01-29  & 01:39:47    &  8751   &   8385  &   348$\pm$22   & 4.974 \\
                 &               &     MOS2 (Th)       &             &             &         &   8386  &   344$\pm$21   &        \\
         10      & 0145740501    &     MOS1 (Th)       & 2003-01-29  & 23:54:24    &  7249   &   6901  &   379$\pm$21   & 4.974  \\
                 &               &     MOS2 (Th)       &             &             &         &   6903  &   351$\pm$23   &        \\
         11      & 0160160101    &     MOS2 (Th)       & 2003-06-08  & 13:29:39    &  38352  &  16870  &   835$\pm$35   & 4.974  \\
         12      & 0160160901    &     MOS2 (Th)       & 2003-06-13  & 23:51:15    &  31655  &  31090  &  1371$\pm$38   & 4.974  \\
         13      & 0145780101    &     MOS2 (Med)      & 2003-07-22  & 01:50:45    &  8736   &   8383  &   510$\pm$26   & 4.974  \\  
         14      & 0160560101    &     MOS2 (Med)      & 2003-08-02  & 21:00:17    &  17952  &  12200  &   722$\pm$29   & 4.974   \\ 
         15      & 0160560201    &     MOS2 (Med)      & 2003-08-09  & 01:43:21    &  12952  &  12060  &   657$\pm$29   & 4.974   \\
         16      & 0160560301    &     MOS2 (Med)      & 2003-08-18  & 15:22:43    &  19143  &  18550  &   945$\pm$37   & 4.974    \\
         17      & 0311990101    &     PN   (Th)       & 2006-01-31  & 18:03:33    &  66949  &  23920  &  3626$\pm$72   & 4.967    \\
         18      & 0560580101    &     MOS2 (Th)       & 2009-01-05  & 10:22:08    &  14916  &  14030  &   512$\pm$24   & 4.933   \\
         19      & 0560580201    &     MOS2 (Th)       & 2009-01-09  & 14:27:16    &  11910  &  11450  &   485$\pm$25   & 4.933    \\
         20      & 0560580301    &     MOS2 (Th)       & 2009-01-15  & 11:22:01    &  26917  &  26140  &  1253$\pm$39   & 4.933    \\
         21      & 0560580401    &     MOS2 (Th)       & 2009-02-02  & 04:45:24    &  26917  &  26230  &  1225$\pm$41   & 4.933    \\
                 &               &      PN  (Th)       &             &             &         &  22770  &  3884$\pm$76     &        \\
         22      & 0650840101    &     MOS1 (Thin)     & 2010-12-06  &  00:08:05   &  90917  &  51170  &   1044$\pm$42   & 12.026    \\
                 &               &     MOS2 (Thin)     &              &            &         &  26770  & 384$\pm$28      &     \\
         23      & 0691970101    &     MOS1 (Thin)     & 2012-12-20   &  19:39:21  &  87714  &  49760  &  1180$\pm$44    & 12.026    \\ 
                 &               &     MOS2 (Thin)     &              &            &         &  72950  &  1572$\pm$54     &     \\         
         24      & 0742850301    &     MOS2 (Med)      & 2014-06-06  & 19:13:05    &  14300  &  12800  &   685$\pm$31   & 4.933    \\
         25      & 0742850401    &     MOS2 (Med)      & 2014-07-28  & 15:32:43    &  35000  &  33320  &  1889$\pm$51   & 4.933    \\
         26      & 0762910401    &     MOS2 (Med)      & 2015-07-16  & 01:18:44    &  13000  &  11500  &   645$\pm$29   & 4.930   \\  
         27      & 0804950201    &     MOS1 (Med)      & 2017-06-04  & 20:47:06    & 33000   &  31210  &    670$\pm$35 &  12.291      \\
                 &               &     MOS2 (Med)      &             &             &         &  31190  &    753$\pm$37 &       \\         
         28      & 0804950301    &     MOS2 (Med)      & 2017-12-06  & 06:26:36    & 30000   &  20140  &   550$\pm$29  &  12.291      \\      
         29      & 0830191801    &     MOS1 (Med)      & 2018-08-22  & 03:14:17    & 33100   &  26770  &   446$\pm$30  &  12.291     \\
                 &               &     MOS2 (Med)      &             &             &         &  29290  &   469$\pm$31 &       \\   
         30      & 0845030201    &     MOS1 (Med)      & 2019-06-07  & 05:25:44    & 34300   &  32390  &    783$\pm$41 &  12.291     \\
                &                &     MOS2 (Med)      &             &             &         &  32430  &  787$\pm$43  &       \\ 
                &                &     PN   (Med)      &             &             &         &  26700  &   2848$\pm$69  &       \\ 
         31      & 0845030301    &     MOS2 (Med)      & 2019-12-07  & 17:23:47    & 28000   &  26250  &  629$\pm$34   &  12.291      \\
                 &               &     PN   (Med)      &             &             &         &  15830  & 1320$\pm$44    &       \\ 
\hline
\hline
\end{tabular}
\tablefoot{(a) `Th' and `Med' stand for the thick and medium optical blocking filters, respectively. \\
(b) Total duration of the observation. \\
(c) LIVETIME is the livetime keyword value after making correction for periods of dead time and high-background (wherever located). \\
(d) Background corrected net source counts have been estimated in 0.3-10.0 keV energy range.\\
(e) Offset between HD\,93205 position at EPIC instruments and the telescope pointing.}   
\end{table*}

\begin{table*}
\centering  
\caption{\label{spec_par}Best fit parameters obtained from spectral fitting of HD\,93205 as observed from \textit{XMM-Newton}-EPIC.}
\setlength{\tabcolsep}{1.2pt}   
\renewcommand{\arraystretch}{1.7}
\begin{tabular}{c c c c c c c c c c c c}   
  \hline\hline
Obs. ID  & $\phi$    & $norm_{1}$        & $norm_{2}$  & N$_{H}^{local}$  &  $F^{obs}_{B}$ & $F^{obs}_{S}$ & $F^{obs}_{H}$ & $F^{ism}_{B}$ & $F^{ism}_{S}$ & $F^{ism}_{H}$ & $\chi^{2}_{\nu} (dof)$ \\ 
 \cline{3-4}                  \cline{6-11}                       
  &           & \multicolumn{2}{c}{($10^{-3}$ cm$^{-5}$)} & ($10^{22}$ cm$^{-2}$)   &   \multicolumn{6}{c}{($10^{-12}$ erg cm$^{-2}$ s$^{-1}$)} &    \\
  \hline
 0112580601& 0.075 &  $2.59^{+0.78}_{-0.60}$  & $0.47^{+0.03}_{-0.03}$ &$0.29^{+0.04}_{-0.03}$  &$0.50^{+0.01}_{-0.01}$ &$ 0.48^{+0.01}_{-0.01}$ &$0.024^{+0.001}_{-0.001}$ &$1.25^{+0.03}_{-0.03}$ &$1.23^{+0.03}_{-0.03}$ &  $0.025^{+0.001}_{-0.001}$ & 1.25 (264)\\
 0112580701& 0.322 &  $3.29^{+1.58}_{-1.16}$  & $0.33^{+0.05}_{-0.06}$ &$0.35^{+0.06}_{-0.07}$  &$0.39^{+0.02}_{-0.02}$ &$ 0.38^{+0.02}_{-0.02}$ &$0.017^{+0.002}_{-0.002}$ &$0.98^{+0.05}_{-0.05}$ &$0.96^{+0.05}_{-0.05}$ &  $0.018^{+0.002}_{-0.002}$ & 1.22 (130)\\
 0112560101& 0.007 &  $3.29^{+0.90}_{-0.63}$  & $0.50^{+0.03}_{-0.03}$ &$0.35^{+0.03}_{-0.03}$  &$0.49^{+0.01}_{-0.01}$ &$ 0.46^{+0.01}_{-0.01}$ &$0.025^{+0.001}_{-0.001}$ &$1.15^{+0.03}_{-0.03}$ &$1.12^{+0.03}_{-0.03}$ &  $0.027^{+0.001}_{-0.001}$ & 1.55 (371)\\
 0112560201& 0.500 &  $2.29^{+0.91}_{-0.67}$  & $0.27^{+0.02}_{-0.03}$ &$0.29^{+0.05}_{-0.05}$  &$0.35^{+0.01}_{-0.01}$ &$ 0.33^{+0.01}_{-0.01}$ &$0.014^{+0.001}_{-0.001}$ &$0.89^{+0.03}_{-0.03}$ &$0.88^{+0.03}_{-0.03}$ &  $0.015^{+0.001}_{-0.001}$ & 1.17 (284)\\
 0112560301& 0.829 &  $2.46^{+0.57}_{-0.49}$  & $0.32^{+0.02}_{-0.02}$ &$0.34^{+0.03}_{-0.03}$  &$0.33^{+0.01}_{-0.01}$ &$ 0.32^{+0.01}_{-0.01}$ &$0.016^{+0.001}_{-0.001}$ &$0.81^{+0.02}_{-0.02}$ &$0.79^{+0.02}_{-0.02}$ &  $0.017^{+0.001}_{-0.001}$ & 1.27 (365)\\
 0145740101& 0.232 &  $1.58^{+1.44}_{-0.81}$  & $0.22^{+0.06}_{-0.05}$ &$0.21^{+0.10}_{-0.10}$  &$0.36^{+0.03}_{-0.03}$ &$ 0.35^{+0.03}_{-0.03}$ &$0.011^{+0.003}_{-0.003}$ &$1.02^{+0.11}_{-0.11}$ &$1.00^{+0.12}_{-0.12}$ &  $0.012^{+0.003}_{-0.003}$ & 1.09 (46)\\
 0145740201& 0.479 &  $2.81^{+2.84}_{-1.53}$  & $0.19^{+0.07}_{-0.10}$ &$0.32^{+0.13}_{-0.12}$  &$0.32^{+0.03}_{-0.03}$ &$ 0.31^{+0.03}_{-0.03}$ &$0.011^{+0.004}_{-0.004}$ &$0.87^{+0.09}_{-0.09}$ &$0.86^{+0.09}_{-0.09}$ &  $0.011^{+0.004}_{-0.004}$ & 0.89 (46)\\
 0145740301& 0.644 &  $2.19^{+2.03}_{-1.16}$  & $0.36^{+0.09}_{-0.08}$ &$0.35^{+0.10}_{-0.10}$  &$0.34^{+0.03}_{-0.03}$ &$ 0.32^{+0.03}_{-0.03}$ &$0.019^{+0.004}_{-0.004}$ &$0.79^{+0.09}_{-0.09}$ &$0.78^{+0.09}_{-0.09}$ &  $0.019^{+0.004}_{-0.004}$ & 0.97 (52)\\
 0145740401& 0.808 &  $3.09^{+2.48}_{-1.33}$  & $0.31^{+0.07}_{-0.09}$ &$0.37^{+0.10}_{-0.09}$  &$0.34^{+0.03}_{-0.03}$ &$ 0.32^{+0.02}_{-0.02}$ &$0.016^{+0.003}_{-0.003}$ &$0.83^{+0.08}_{-0.08}$ &$0.81^{+0.08}_{-0.08}$ &  $0.017^{+0.004}_{-0.004}$ & 1.46 (64)\\
 0145740501& 0.973 &  $4.06^{+2.94}_{-1.95}$  & $0.43^{+0.09}_{-0.10}$ &$0.38^{+0.09}_{-0.09}$  &$0.44^{+0.03}_{-0.03}$ &$ 0.42^{+0.03}_{-0.03}$ &$0.022^{+0.005}_{-0.005}$ &$1.05^{+0.09}_{-0.09}$ &$1.03^{+0.09}_{-0.09}$ &  $0.023^{+0.005}_{-0.005}$ & 1.14 (65)\\
 0160160101& 0.312 &  $2.89^{+2.22}_{-1.39}$  & $0.25^{+0.06}_{-0.08}$ &$0.31^{+0.09}_{-0.09}$  &$0.37^{+0.03}_{-0.03}$ &$ 0.36^{+0.03}_{-0.03}$ &$0.013^{+0.003}_{-0.003}$ &$0.98^{+0.09}_{-0.09}$ &$0.97^{+0.09}_{-0.09}$ &  $0.014^{+0.003}_{-0.003}$ & 1.33 (67)\\
 0160160901& 0.217 &  $2.50^{+1.43}_{-1.02}$  & $0.32^{+0.04}_{-0.05}$ &$0.31^{+0.07}_{-0.07}$  &$0.38^{+0.02}_{-0.02}$ &$ 0.36^{+0.02}_{-0.02}$ &$0.016^{+0.002}_{-0.002}$ &$0.96^{+0.06}_{-0.06}$ &$0.95^{+0.06}_{-0.06}$ &  $0.017^{+0.002}_{-0.002}$ & 1.38 (88)\\
 0145780101& 0.466 &  $2.06^{+1.98}_{-0.10}$  & $0.36^{+0.09}_{-0.09}$ &$0.29^{+0.11}_{-0.11}$  &$0.39^{+0.03}_{-0.03}$ &$ 0.38^{+0.03}_{-0.03}$ &$0.019^{+0.004}_{-0.004}$ &$0.99^{+0.11}_{-0.11}$ &$0.97^{+0.11}_{-0.11}$ &  $0.019^{+0.004}_{-0.004}$ & 1.32 (44)\\
 0160560101& 0.399 &  $2.86^{+2.59}_{-1.49}$  & $0.36^{+0.07}_{-0.08}$ &$0.33^{+0.10}_{-0.10}$  &$0.39^{+0.03}_{-0.03}$ &$ 0.38^{+0.03}_{-0.03}$ &$0.018^{+0.003}_{-0.003}$ &$0.99^{+0.09}_{-0.09}$ &$0.97^{+0.09}_{-0.09}$ &  $0.019^{+0.004}_{-0.004}$ & 1.26 (58)\\
 0160560201& 0.427 &  $1.55^{+1.88}_{-0.85}$  & $0.25^{+0.05}_{-0.05}$ &$0.21^{+0.12}_{-0.10}$  &$0.38^{+0.03}_{-0.03}$ &$ 0.36^{+0.03}_{-0.03}$ &$0.013^{+0.002}_{-0.002}$ &$1.06^{+0.09}_{-0.09}$ &$1.04^{+0.09}_{-0.09}$ &  $0.014^{+0.002}_{-0.002}$ & 1.19 (54)\\
 0160560301& 0.989 &  $3.41^{+2.48}_{-1.68}$  & $0.31^{+0.07}_{-0.08}$ &$0.38^{+0.09}_{-0.09}$  &$0.35^{+0.02}_{-0.02}$ &$ 0.34^{+0.02}_{-0.02}$ &$0.016^{+0.003}_{-0.003}$ &$0.86^{+0.06}_{-0.06}$ &$0.84^{+0.06}_{-0.06}$ &  $0.017^{+0.003}_{-0.003}$ & 1.17 (74)\\
 0311990101& 0.597 &  $2.52^{+0.98}_{-0.74}$  & $0.34^{+0.03}_{-0.03}$ &$0.33^{+0.05}_{-0.05}$  &$0.37^{+0.01}_{-0.01}$ &$ 0.35^{+0.01}_{-0.01}$ &$0.017^{+0.001}_{-0.001}$ &$0.91^{+0.03}_{-0.03}$ &$0.89^{+0.03}_{-0.03}$ &  $0.018^{+0.002}_{-0.002}$ & 1.32 (146)\\
 0560580101& 0.452 &  $2.35^{+2.23}_{-1.26}$  & $0.30^{+0.07}_{-0.07}$ &$0.33^{+0.11}_{-0.11}$  &$0.33^{+0.03}_{-0.03}$ &$ 0.32^{+0.03}_{-0.03}$ &$0.015^{+0.003}_{-0.003}$ &$0.82^{+0.08}_{-0.08}$ &$0.81^{+0.08}_{-0.08}$ &  $0.016^{+0.004}_{-0.004}$ & 1.41 (44)\\
 0560580201& 0.151 &  $6.28^{+5.12}_{-3.56}$  & $0.41^{+0.15}_{-0.18}$ &$0.46^{+0.11}_{-0.12}$  &$0.44^{+0.04}_{-0.04}$ &$ 0.42^{+0.04}_{-0.04}$ &$0.021^{+0.007}_{-0.007}$ &$1.02^{+0.10}_{-0.10}$ &$0.99^{+0.10}_{-0.10}$ &  $0.022^{+0.007}_{-0.007}$ & 0.95 (43)\\
 0560580301& 0.138 &  $5.05^{+2.53}_{-2.07}$  & $0.34^{+0.07}_{-0.08}$ &$0.38^{+0.07}_{-0.08}$  &$0.46^{+0.02}_{-0.02}$ &$ 0.44^{+0.02}_{-0.02}$ &$0.018^{+0.003}_{-0.003}$ &$1.15^{+0.07}_{-0.07}$ &$1.13^{+0.07}_{-0.07}$ &  $0.019^{+0.003}_{-0.003}$ & 1.11 (87)\\
 0560580401& 0.016 &  $2.84^{+0.79}_{-0.72}$  & $0.36^{+0.03}_{-0.03}$ &$0.31^{+0.04}_{-0.04}$  &$0.42^{+0.01}_{-0.01}$ &$ 0.40^{+0.01}_{-0.01}$ &$0.018^{+0.001}_{-0.001}$ &$1.05^{+0.03}_{-0.03}$ &$1.04^{+0.03}_{-0.03}$ &  $0.019^{+0.001}_{-0.001}$ & 1.19 (237)\\
 0650840101& 0.578 &  $3.34^{+1.99}_{-1.45}$  & $0.33^{+0.06}_{-0.07}$ &$0.39^{+0.08}_{-0.08}$  &$0.34^{+0.02}_{-0.02}$ &$ 0.32^{+0.02}_{-0.02}$ &$0.017^{+0.003}_{-0.003}$ &$0.79^{+0.05}_{-0.05}$ &$0.78^{+0.05}_{-0.05}$ &  $0.018^{+0.003}_{-0.003}$ & 1.20 (128)\\
 0691970101& 0.228 &  $3.78^{+2.21}_{-1.57}$  & $0.49^{+0.05}_{-0.07}$ &$0.43^{+0.07}_{-0.07}$  &$0.39^{+0.02}_{-0.02}$ &$ 0.37^{+0.01}_{-0.01}$ &$0.024^{+0.003}_{-0.003}$ &$0.88^{+0.04}_{-0.04}$ &$0.85^{+0.04}_{-0.04}$ &  $0.026^{+0.003}_{-0.003}$ & 1.32 (204)\\
 0742850301& 0.847 &  $2.54^{+1.96}_{-1.22}$  & $0.30^{+0.07}_{-0.07}$ &$0.33^{+0.09}_{-0.09}$  &$0.34^{+0.03}_{-0.03}$ &$ 0.33^{+0.03}_{-0.03}$ &$0.016^{+0.003}_{-0.003}$ &$0.85^{+0.08}_{-0.08}$ &$0.83^{+0.08}_{-0.08}$ &  $0.017^{+0.004}_{-0.004}$ & 1.39 (57)\\
 0742850401& 0.358 &  $2.30^{+1.26}_{-0.87}$  & $0.30^{+0.04}_{-0.04}$ &$0.27^{+0.07}_{-0.07}$  &$0.40^{+0.02}_{-0.02}$ &$ 0.39^{+0.02}_{-0.02}$ &$0.016^{+0.002}_{-0.002}$ &$1.06^{+0.06}_{-0.06}$ &$1.05^{+0.06}_{-0.06}$ &  $0.017^{+0.002}_{-0.002}$ & 1.13 (109)\\
 0762910401& 0.332 &  $2.29^{+2.01}_{-1.09}$  & $0.34^{+0.06}_{-0.07}$ &$0.29^{+0.09}_{-0.09}$  &$0.39^{+0.03}_{-0.03}$ &$ 0.37^{+0.03}_{-0.03}$ &$0.018^{+0.003}_{-0.003}$ &$0.97^{+0.09}_{-0.09}$ &$0.95^{+0.09}_{-0.09}$ &  $0.019^{+0.003}_{-0.003}$ & 1.17 (56)\\
 0804950201& 0.772 &  $2.20^{+1.27}_{-0.92}$  & $0.29^{+0.05}_{-0.05}$ &$0.29^{+0.07}_{-0.07}$  &$0.36^{+0.02}_{-0.02}$ &$ 0.34^{+0.02}_{-0.02}$ &$0.015^{+0.002}_{-0.002}$ &$0.92^{+0.06}_{-0.06}$ &$0.90^{+0.06}_{-0.06}$ &  $0.016^{+0.002}_{-0.002}$ & 1.29 (129)\\
 0804950301& 0.116 &  $5.74^{+4.82}_{-2.84}$  & $0.48^{+0.14}_{-0.17}$ &$0.47^{+0.11}_{-0.11}$  &$0.43^{+0.04}_{-0.04}$ &$ 0.41^{+0.03}_{-0.03}$ &$0.024^{+0.007}_{-0.007}$ &$0.97^{+0.09}_{-0.09}$ &$0.95^{+0.09}_{-0.09}$ &  $0.026^{+0.007}_{-0.007}$ & 0.96 (49)\\
 0830191801& 0.672 &  $1.89^{+1.79}_{-0.98}$  & $0.24^{+0.05}_{-0.06}$ &$0.29^{+0.10}_{-0.10}$  &$0.30^{+0.02}_{-0.02}$ &$ 0.29^{+0.02}_{-0.02}$ &$0.012^{+0.002}_{-0.002}$ &$0.78^{+0.07}_{-0.07}$ &$0.76^{+0.07}_{-0.07}$ &  $0.013^{+0.002}_{-0.003}$ & 1.21 (91)\\
 0845030201& 0.243 &  $1.92^{+0.81}_{-0.58}$  & $0.39^{+0.03}_{-0.03}$ &$0.26^{+0.05}_{-0.05}$  &$0.44^{+0.01}_{-0.01}$ &$ 0.42^{+0.01}_{-0.01}$ &$0.020^{+0.001}_{-0.001}$ &$1.10^{+0.04}_{-0.04}$ &$1.08^{+0.04}_{-0.04}$ &  $0.021^{+0.001}_{-0.001}$ & 1.26 (280)\\
 0845030301& 0.423 &  $1.76^{+0.94}_{-0.63}$  & $0.33^{+0.04}_{-0.04}$ &$0.28^{+0.06}_{-0.06}$  &$0.36^{+0.02}_{-0.02}$ &$ 0.34^{+0.02}_{-0.02}$ &$0.017^{+0.002}_{-0.002}$ &$0.91^{+0.05}_{-0.05}$ &$0.89^{+0.05}_{-0.05}$ &  $0.018^{+0.002}_{-0.002}$ & 1.18 (150)\\
\hline
\end{tabular}
\tablefoot{Fit parameters are derived from joint spectral fitting of \textit{XMM-Newton}$-$PN, MOS1 and MOS2 spectra of HD\,93205 using model \textsc{phabs(ism)*phabs(local)*(apec+apec)} with fixed values of $N_{H}^{ISM}=0.24 \times 10^{22}$ $cm^{-2}$ and $k$T$_{1}$ = 0.2 keV and $k$T$_{2}$ = 0.6 keV. $norm_{1}$ and $norm_{2}$ are the normalization parameters for two temperature components whereas $N_{H}^{local}$ is the equivalent local H-column density. $F_{B}^{obs}$,  $F_{S}^{obs}$, and $F_{H}^{obs}$ are the observed and $F_{B}^{ism}$,  $F_{S}^{ism}$, and $F_{H}^{ism}$ are the ISM corrected X-ray fluxes of HD\,93205 in broad, soft, and hard energy bands, respectively. $\chi_{\nu}^{2}$ is the reduced $\chi^2$  and  \textit{dof} is degrees of freedom.  Errors quoted on different parameters refer to 90\% confidence level.}
\end{table*}

\end{appendix}

\end{document}